\documentclass[reviewcopy]{elsarticle}

\usepackage[reviewcopy]{adndt}
\usepackage{longtable}
\usepackage{url}
\RequirePackage{subfigure}
\usepackage{amsmath}
\usepackage{amssymb}

\newcommand{\etal} {{\it et al}}
\newcommand{\VS} {\vspace{0.2cm}}
\newcommand{\LS} {$LS$}
\newcommand{\Rm} {{\bf{R}}}
\newcommand{\SLP} [3]{$^{#1}$#2$^{\rm{#3}}$}
\newcommand{\SLPJ}[4]{$^{#1}$#2$^{\rm{#3}}_{_{#4}}$}
\newcommand{\nlo} [3]{#1#2$^{#3}$}
\newcommand{\po}  [2]{\overline{#1}\rm{#2}}
\newcommand{\AS} {Autostructure}
\newcommand{\pcs} {photoionization cross section}
\newcommand{\STOCi} {C_{i}}
\newcommand{\STOPi} {P_{i}}
\newcommand{\STOxii} {\xi_{i}}
\newcommand{\Par} {\pi}
\newcommand{\QB} {{\bf{QB}}}
\newcommand{\Sm} {{\bf{S}}}
\newcommand{\Km} {{\bf{K}}}
\newcommand{\Mm} {{\bf{\M}}}
\newcommand{\M} {M}
\newcommand{\iu} {\textrm{i}}
\newcommand{\Dirac} {\hbar}
\newcommand{\E} {E}
\newcommand{\Eo} {\E_{0}}
\newcommand{\BGCK} {K_{_{o}}}
\newcommand{\PKF} {{\rm{g}}}
\newcommand{\Ei} {\E_{i}}
\newcommand{\SVKi} {K_{i}}
\newcommand{\Er} {\E_{r}}
\newcommand{\FWHMe}  {\Delta_{E}}
\newcommand{\RW} {\Delta_{r}}
\newcommand{\OS} {f}
\newcommand{\WL} {\lambda}
\newcommand{\cC} [1] {\multicolumn{1}{c}{#1}}
\newcommand{\EE} [1] {$^{^{#1}}$}
\newcommand{\II}      {~{\sc ii}}
\newcommand{\III}     {~{\sc iii}}
\newcommand{\IV}      {~{\sc iv}}
\newcommand{\ND}     {N}
\newcommand{\NDe}    {\ND_{e}}
\newcommand{\NDi}    {\ND_{i}}
\newcommand{\NDl}    {\ND_{l}}
\newcommand{\NDu}    {\ND_{u}}
\newcommand{\RC}     {\varrho}
\newcommand{\RCc}    {\RC_{_{c}}}
\newcommand{\RCf}    {\RC_{_{f}}}
\newcommand{\rTP}    {\TP^{r}}
\newcommand{\rTPl}   {\rTP_{l}}
\newcommand{\rTPul}  {\rTP_{ul}}
\newcommand{\aTP}    {\TP^{a}}
\newcommand{\TP}     {\Gamma}
\newcommand{\aTPl}   {\aTP_{l}}
\newcommand{\h}      {h}
\newcommand{\F}      {\nu}
\newcommand{\EMISS}  {\varepsilon}
\newcommand{\ENLul}  {\EMISS_{ul}}

\biboptions{square,sort&compress}
\bibpunct[]{[}{]}{,}{n}{}{;}
\begin{document}

\begin{frontmatter}

\journal{Atomic Data and Nuclear Data Tables}

\title{Dielectronic Recombination Lines of C$^+$}

  \author[One]{Taha Sochi\fnref{X}\corref{cor1}}
  \ead{E-mail: t.sochi@ucl.ac.uk}

  \author[One,Two]{Peter J. Storey\fnref{Y}}


  \cortext[cor1]{Corresponding author.}

  \address[One]{University College London, Department of Physics and Astronomy, Gower Street, London, WC1E 6BT}


\date{16.12.2002} 

\begin{abstract}
The current paper presents atomic data generated to investigate the recombination lines of C~{\sc
ii} in the spectra of planetary nebulae. These data include energies of bound and autoionizing
states, oscillator strengths and radiative transition probabilities, autoionization probabilities,
and recombination coefficients. The R-matrix method of electron scattering theory was used to
describe the C$^{2+}$ plus electron system.

\end{abstract}

\end{frontmatter}


\newpage

\tableofcontents
\listofDtables
\listofDfigures
\vskip5pc

\newpage

\section{Introduction}

The study of atomic and ionic lines of carbon in the spectra of astronomical objects, such as
planetary nebulae \cite{PottaschWD1978, NikitinSFK1981, BogdanovichLNRK1985}, has implications for
carbon abundance determination \cite{HarringtonLSS1980, HarringtonLS1981, NussbaumerS19812,
AdamsS1982, Kholtygin1984, BogdanovichNRK1985}, probing the physical conditions in the interstellar
medium \cite{Boughton1978, PengHWL2005}, and element enrichment in the CNO cycle \cite{AdamsS1982,
RolaS1994, CleggSWN1997}. The lines span large parts of the electromagnetic spectrum and originate
from various processes under different physical conditions. Concerning C$^+$, the subject of the
current investigation, spectral lines have been observed in many astronomical objects such as
planetary nebulae, Seyfert galaxies, stellar winds of Wolf-Rayet stars, symbiotic stars, and in the
interstellar medium \cite{NikitinY1963, HayesN1984, Kholtygin1984, RolaS1994, DaveySK2000,
PengWHL2004, ZhangLLPB2005, StanghelliniSG2005}.

Many atomic processes can contribute to the production of carbon spectral lines, including
radiative and collisional excitation, and radiative and dielectronic recombination. The current
study is concerned with some of the lines produced by dielectronic recombination and subsequent
radiative cascades.

Capture of an incoming electron to a target ion may occur through a non-resonant background
continuum, which is radiative recombination (RR), or through a process involving doubly-excited
states (resonances) known as dielectronic recombination (DR). The latter can lead either to
autoionization, a radiationless transition to a lower state of the ion with the ejection of a free
electron, or to stabilization by radiative decay to a lower true bound state.

The rate for RR rises as the free electron temperature falls and hence tends to be the dominant
recombination process at low temperatures. It is particularly significant in the cold ionized
plasmas found, for example, in some supernova remnants \cite{PequignotPB1991, NaharPZ2000,
BadnellOSABe2003}.

We may distinguish three types of DR mechanism with relevance at different temperatures

\begin{enumerate}

\item
High-temperature DR (HTDR) which occurs through Rydberg series of autoionizing states, and in which
the radiative stabilization is via a decay within the ion core.

\item
Low-temperature DR (LTDR) which operates via a few near-threshold resonances with radiative
stabilization usually through decay of the outer, captured electron. These resonances are usually
within a few thousand wave numbers of threshold, so this process operates at thousands to tens of
thousands of degrees Kelvin.

\item
Fine-structure DR (FSDR) which is due to Rydberg series resonances converging on the fine-structure
levels of the ground term of the recombining ion, and is necessarily stabilized by outer electron
decays. This process can operate at very low temperatures, down to tens or hundreds of Kelvin.

\end{enumerate}
In this paper we are concerned only with LTDR and the resulting spectral lines.

With regard to the recombination lines of C~{\sc ii}, there are many theoretical and observational
studies mainly within astronomical contexts. Examples are \cite{Leibowitz1972, Leibowitz1972b,
BalickGD1974, NussbaumerS1975, Boughton1978, ClavelFS1981, HarringtonLS1981, NussbaumerS19812,
CleggSPP1983, HayesN1984, NussbaumerS1984, BogdanovichNRK1985, YanS1987, Badnell1988,
PequignotPB1991, RolaS1994, BaluteauZMP1995, Kholtygin1998, DaveySK2000, LiuLBL2004, PengWHL2004,
PengHWL2005, ZhangLLPB2005}. From the perspective of atomic physics, the most comprehensive of
these studies and most relevant to the current investigation is that of Badnell \cite{Badnell1988}
who calculated configuration mixing effective dielectronic recombination coefficients for the
recombined C$^+$ ion at temperatures $T = 10^{3}-6.3\times10^{4}$\,K applicable to planetary
nebulae, and Davey \etal\ \cite{DaveySK2000} who computed effective recombination coefficients for
C~{\sc ii} transitions between doublet states for the temperature range 500-20000\,K with an
electron density of 10$^{4}$ cm$^{-3}$ relevant to planetary nebulae. Badnell performed
calculations in both \LS\ and intermediate coupling schemes and over a wide temperature range,
using the \AS\ code which treats the autoionizing states as bound and the interaction with
continuum states by a perturbative approach. On the other hand, Davey \etal\ performed their
calculations using the \Rm-matrix code which uses a unified treatment of bound and continuum states
but worked in the \LS-coupling scheme and hence their results were limited to doublet states.

The aim of the current study is to treat the near-threshold resonances and subsequent radiative
decays using the unified approach of the \Rm-matrix method in intermediate coupling. The
investigation includes all the autoionizing resonance states above the threshold of C$^{2+}$
\SLP1Se with a principal quantum number $n<5$ for the captured electron as an upper limit. This
condition was adopted mainly due to computational limitations. In total, 61 autoionizing states (27
doublets and 34 quartets) with this condition have been theoretically found by the \Rm-matrix
method. Of these 61 resonances, 55 are experimentally observed according to the NIST database
\cite{NIST2010}. More details will follow in the forthcoming sections.

\section{Tools and Methods of Calculation}

\subsection{Autoionizing and Bound States Calculations}\label{AutBouCal}

We use the \Rm-matrix code \cite{BerringtonEN1995}, and \AS\ \cite{EissnerJN1974, NussbaumerS1978,
BadnellAS2008} to compute the properties of autoionizing and bound states of C$^+$. Orbitals
describing the target for the \Rm-matrix scattering calculations were taken from Berrington \etal\
\cite{BerringtonBDK1977}, who used a  target comprising the six terms \nlo2s2 \SLP1Se,
\nlo2s{}\nlo2p{} \SLP3Po, \SLP1Po and \nlo2p2 \SLP3Pe, \SLP1De, \SLP1Se, constructed from seven
orthogonal orbitals; three physical and four pseudo orbitals. These orbitals are: 1s, 2s, 2p,
$\po3s$, $\po3p$, $\po3d$ and $\po4f$, where the bar denotes a pseudo orbital. The purpose of
including pseudo orbitals is to represent electron correlation effects and to improve the target
wavefunctions. The radial parts, $P_{nl}(r)$, of these orbitals are Slater Type Orbital (STO)
generated by the CIV3 program of Hibbert \cite{Hibbert1975}. Each STO is defined by
\begin{equation}\label{STO}
    P_{nl}(r) = \sum_{i} \STOCi r^{\STOPi} e^{- \STOxii r}
\end{equation}
where $\STOCi$ is a coefficient and $\STOPi$ and $\STOxii$ are indicial parameters in this
specification, $r$ is the radius, and $i$ is a counting index that runs over the orbitals of
interest. The values of these parameters are given in Table (\ref{SevenorbitalsSTO}).

In this work we construct a scattering target of 26 terms which includes the 6 terms of Berrington
\etal\ \cite{BerringtonBDK1977} plus all terms of the configurations \nlo2s{}$\po3l$ and
\nlo2p{}$\po3l$, l$=0,1,2$.  This includes the terms outside the $n=2$ complex which make the
largest contribution to the dipole polarizability of the \nlo2s2 \SLP1Se and \nlo2s{}\nlo2p{}
\SLP3Po states of C$^{2+}$.

The \Rm-matrix calculations were carried out in the intermediate coupling (IC) scheme by including
the spin-orbit interaction terms of the Breit-Pauli Hamiltonian. The requirement for the IC scheme
arises from the fact that in \LS-coupling the conserved quantities are $LS\Par$ and hence only the
doublet states that conserve these quantities, such as \SLP2Se and \SLP2Po, can autoionize.
Therefore, in \LS-coupling no autoionization is allowed for quartet terms or some doublet states,
such as \SLP2So and \SLP2Pe.

The number of continuum basis orbitals used to express the wavefunction in the inner region (MAXC
in STG1 of the \Rm-matrix code) was varied between 6-41 and the results were analyzed. It was
noticed that increasing the number of basis functions, with all ensuing computational costs, does
not necessarily improve the results; moreover a convergence instability may occur in some cases. It
was decided therefore to use MAXC = 16 in all calculations as a compromise between the
computational resources requirement and accuracy. The effect of varying the size of the inner
region radius in the \Rm-matrix formulation was also investigated and a value of 10 atomic units
was chosen on the basis of numerical stability and convergence.

Several methods exist for finding and analyzing resonances that arise during the recombination
processes; these methods include the Time-Delay method of Stibbe and Tennyson \cite{StibbeT1998}
which is based on the use of lifetime matrix eigenvalues \cite{Smith1960} to locate the resonance
position and identify its width, and the \QB\ method of Quigley \etal\ \cite{QuigleyB1996,
QuigleyBP1998} which applies a fitting procedure to the reactance matrix eigenphase near the
resonance position using the analytic properties of the \Rm-matrix theory.

For the low-lying resonances just above the \SLP1Se\ ionization threshold, the scattering matrix
\Sm\ has only one channel, and hence the reactance matrix, \Km, is a real scalar with a pole near
the resonance position. According to the collision theory of Smith \cite{Smith1960}, the lifetime
matrix \Mm\ is related to the \Sm-matrix by

\begin{equation}\label{TDmatrixS}
    \Mm = -\iu \, \Dirac \, \Sm^{*} \frac{d \Sm}{d \E}
\end{equation}
where \iu\ is the imaginary unit, $\Dirac$ ($=h/2\pi$) is the reduced Planck's constant, $\Sm^{*}$
is the complex conjugate of $\Sm$, and $\E$ is the energy. Now, a \Km-matrix with a pole at energy
$\Eo$ superimposed on a background $\BGCK$ can be approximated by
\begin{equation}\label{Kmatrix}
    \SVKi = \BGCK + \frac{\PKF}{\Ei - \Eo}
\end{equation}
where $\SVKi$ is the value of \Km-matrix at energy $\Ei$ and $\PKF$ is a physical parameter with
dimension of energy. It follows that in the case of single-channel scattering the \Mm-matrix is
real with a value given by
\begin{equation}\label{mmatrix}
    \M = \frac{-2 \PKF}{(1+ \BGCK^{2})(\E-\Eo)^{2}+2 \BGCK \PKF (\E-\Eo) + \PKF^{2}}
\end{equation}
Using the fact demonstrated by Smith \cite{Smith1960} that the lifetime of the state is the
expectation value of $\M$, it can be shown from Equation (\ref{mmatrix}) that the position of the
resonance peak $\Er$ is given by
\begin{equation}\label{Er}
    \Er = \Eo - \frac{\BGCK \PKF}{1 + \BGCK^{2}}
\end{equation}
while the full width at half maximum $\FWHMe$ is given by
\begin{equation}\label{FWHM}
    \FWHMe = \frac{|2 \PKF|}{1 + \BGCK^{2}}
\end{equation}

The two parameters of primary interest to our investigation are the resonance energy position
$\Er$, and the resonance width $\RW$ which equals the full width at half maximum $\FWHMe$. However,
for an energy point $\Ei$ with a \Km-matrix value $\SVKi$, Equation (\ref{Kmatrix}) has three
unknowns, $\BGCK$, $\PKF$ and $\Eo$, which are needed to find $\Er$ and $\RW$. Hence, three energy
points in the immediate neighborhood of $\Eo$ are required to identify these unknowns. As the
\Km-matrix changes sign at the pole, the neighborhood of $\Eo$ is located by testing the \Km-matrix
value at each point of the energy mesh for sign change. Consequently, the three points are obtained
and can be used to find $\Er$ and $\RW$.

To improve the performance of this approach, an interactive graphical technique was developed to
read and plot the \Km-matrix data directly while searching for poles. In a later stage, more
efficient non-graphical tools for pole searching were used. The purpose of these tools is to search
for any sudden increase or decrease in the background of the \Km-matrix where poles do exist and
where a search with a finer energy mesh would then be carried out.

With regard to sampling the three points for the \Km-matrix calculations, it was observed that
sampling the points very close to the pole makes the energy position and width of resonances
susceptible to fluctuations and instabilities. Therefore, a sampling scheme was adopted in which
the points are selected from a broad range not too close to the pole. This approach was implemented
by generating two meshes, coarse and fine, around the pole as soon as the pole is found. To check
the results, several different three-point combinations for each resonance were used to find the
position and width of the resonance. In each case, the results from these different combinations
were compared. In all cases they were identical within acceptable numerical errors. The results of
\QB\ confirm this conclusion as they agree with the \Km-matrix results as can be seen in Table
(\ref{RTableKQ}).

The results for all bound and autoionizing states  are given in Tables
(\ref{BTable}-\ref{RTableKQ}). In total, 142 bound states belonging to 11 symmetries ($2J=1,3,5,7,9$
even and $2J=1,3,5,7,9,11$ odd) and 61 resonances belonging to 11 symmetries ($2J=1,3,5,7,9,11$
even and $2J= 1,3,5,7,9$ odd) were identified. As seen in Tables (\ref{BTable}-\ref{RTableKQ}), the
theoretical results for both bound and resonance states agree very well with the available
experimental data both in energy levels and in fine structure splitting. Experimental energies are
not available for the very broad resonances as they are difficult to find experimentally. The
maximum discrepancy between experiment and theory in the worst case does not exceed a few percent.
Furthermore, the ordering of the energy levels is the same between the theoretical and experimental
in most cases. Order reversal in some cases is indicated by a minus sign in the fine structure
splitting.

\subsection{Oscillator Strength Calculations} \label{fValues}

The oscillator strengths for free-free, free-bound and bound-bound transitions are required to find
the radiative probabilities for these transitions. As there is no free-free stage in the available
\Rm-matrix code, the $\OS$-values for the free-free transitions could not be produced by
\Rm-matrix. Therefore, these values were generated by \AS\ in the intermediate coupling scheme
where 60 electron configurations were included in the \AS\ input: 2s$^{2}$ $nl$ (2p$\leq nl
\leq$7s), 2s2p $nl$ (2p$\leq nl \leq$7s), 2p$^{3}$, and 2p$^{2}$ $nl$ (3s$\leq nl \leq$7s). An
iterative procedure was followed to find the orbital scaling parameters ($\lambda$'s) required in
\AS. These parameters are given in Table (\ref{lambdaTable}). The scaling parameters, which are
obtained by Autostructure in an automated optimization variational process, are required to
minimize the weighted energy sum of the included target states \cite{EissnerN1969, EissnerJN1974,
Storey1981}.

\begin{figure}
\centering %
\subfigure[2s2p(\SLP3Po)3d \SLPJ4Po{1/2} resonance in 2s2p$^2$ \SLPJ4Pe{1/2} cross-section.]%
{\begin{minipage}[b]{1\textwidth} \centering \includegraphics[width=3.2in] {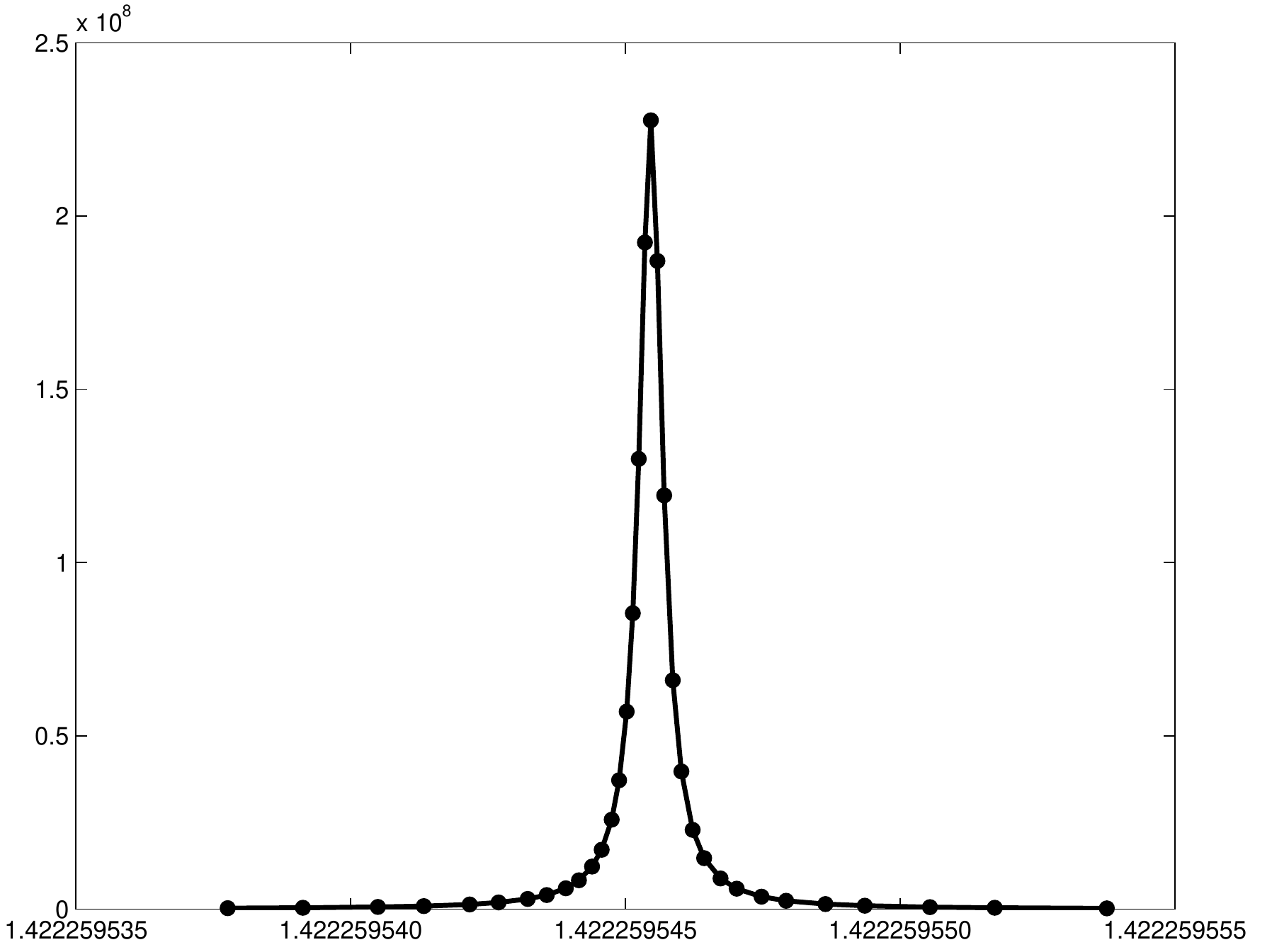}
\end{minipage}} \vspace{-0.1cm}
\centering %
\subfigure[2s2p(\SLP3Po)4d \SLPJ2Fo{5/2} resonance in 2s2p$^2$ \SLPJ4Pe{3/2} cross-section.]%
{\begin{minipage}[b]{1\textwidth} \centering \includegraphics[width=3.2in] {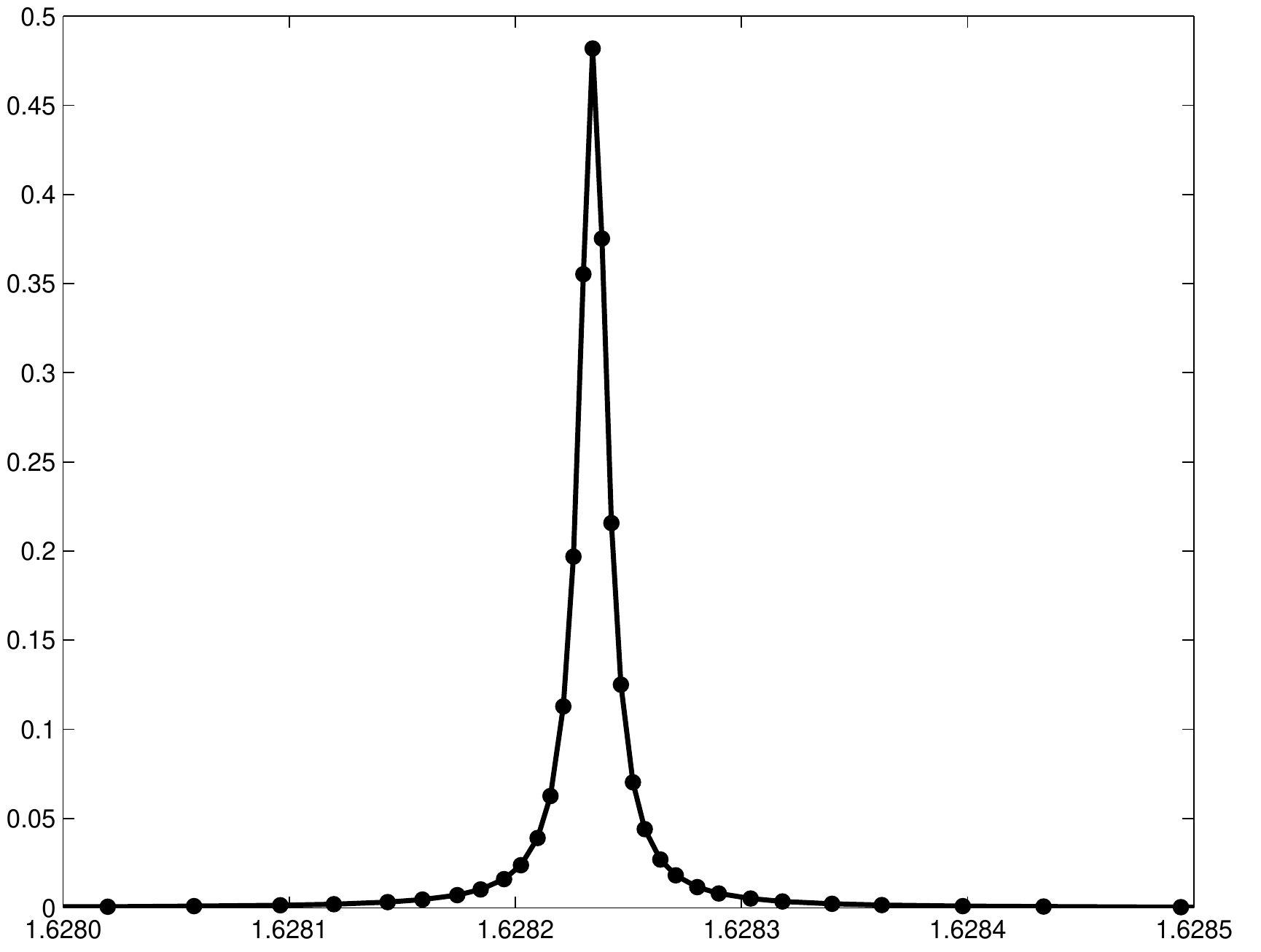}
\end{minipage}} \vspace{-0.1cm}
\centering %
\subfigure[2s2p(\SLP3Po)4p \SLPJ2Pe{3/2} resonance in  2p$^3$ \SLPJ2Do{5/2} cross-section.]%
{\begin{minipage}[b]{1\textwidth} \centering \includegraphics[width=3.2in] {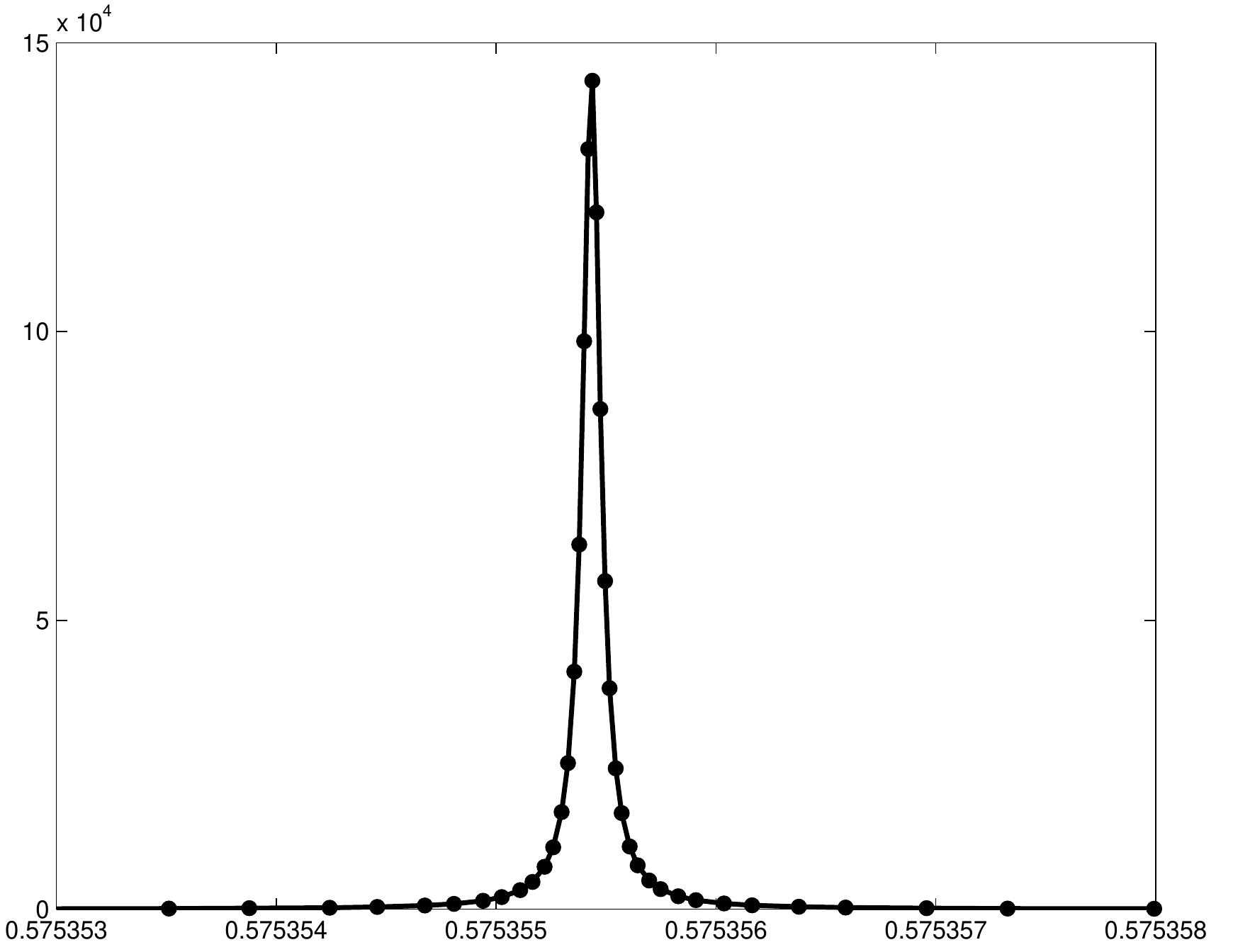}
\end{minipage}}
\caption{Examples of \pcs s in mega barn ($y$-axis) of the indicated bound states near to the
designated autoionizing states versus photon energy in Rydberg ($x$-axis). \label{GGs}}
\end{figure}

Regarding the free-bound transitions, the $\OS$-values for more than 2500 free-bound transitions
were computed by integrating the \pcs s (in mega barn) over the photon energy (in Rydberg). This
was done for each bound state and for all resonances in the corresponding cross-section. The area
under the cross-section curve comprises a background contribution, assumed linear with energy, and
the contribution due to the resonance, which is directly related to the bound-free oscillator
strength. The \pcs s as a function of photon energy were generated by stage STGBF of the \Rm-matrix
code. Some representative cross-sections are shown in Figure (\ref{GGs}) which displays a number of
examples of \pcs s of the indicated bound states close to the designated autoionizing states. It
should be remarked that a separate energy mesh was used to generate \pcs\ data for each individual
resonance with a refinement process to ensure correct mapping and to avoid peak overlapping from
different resonances.

The $\OS$-values of the bound-bound transitions were calculated using stage STGBB of the \Rm-matrix
code. However, the bound-bound $\OS$-values for the 8 uppermost bound states, namely the
2s2p(\SLP3Po)3d~\SLP4Fo and \SLP4Do levels, were generated by \AS\ using similar input data and
procedures to those used in generating the $\OS$-values for the free-free transitions, as outlined
earlier. These states have very large effective quantum number and hence are out of validity range
of the \Rm-matrix code. \AS\ was also used to generate the free-bound $\OS$-values for these 8
uppermost bound states for the same reason.

\subsection{Emissivity Calculations} \label{EmiCal}

We are concerned with spectral lines formed by dielectronic capture followed by radiative decay.
Including only these two processes and autoionization, the number density, $\NDl$ of a state $l$ is
given by
\begin{equation}\label{detbal}
    \NDe \NDi \RCc + \sum_{u} \NDu \rTPul = \NDl (\aTPl + \rTPl)
\end{equation}
where $\RCc$ is the rate coefficient for dielectronic capture to state $l$ and $\aTPl$ is the
autoionization probability of that state, related by
\begin{equation}
\RCc = \left( \frac{\NDl}{\NDe \NDi} \right)_S \aTPl
\end{equation}
where the subscript $S$ refers to the value of the ratio given by the Saha equation, and $\NDe$ and
$\NDi$ are the number density of electrons and ions respectively. If state $l$ lies below the
ionization limit, $\RCc=\aTPl=0$. Equation~\ref{detbal} can be solved for the populations $l$ by a
stepwise downward iteration using the `Emissivity' code \cite{SochiEmis2010}. Note that all other
processes have been neglected, so that the results obtained here are incomplete for states likely
to be populated by radiative recombination or collisional excitation and de-excitation. The results
for free-free and free-bound transitions can be used directly to predict line intensities from
low-density astrophysical plasmas such as gaseous nebulae but those between bound states
underestimate the line intensities in general and should only be used as part of a larger ion
population model including all relevant processes.

The emissivity in transition $u \rightarrow l$ is given by
\begin{equation}\label{emissivity1}
    \ENLul = \NDu \rTPul \h \F
\end{equation}
where $\NDu$ is the population of the upper state, $\rTPul$ is the radiative transition probability
between the upper and lower states, $\h$ is the Planck's constant, and $\F$ is the frequency of the
transition line. The equivalent effective recombination coefficient $\RCf$, which is linked to the
emissivity by the following relation, can also be computed
\begin{equation}\label{EmissRecCoeff}
    \RCf = \frac{\EMISS} {\NDe \NDi \h \F}
\end{equation}
where $\NDe$ and $\NDi$ are the number density of electrons and ions respectively. In these
calculations, all theoretical data for the energy of resonances and bound states were replaced with
experimental data from NIST when such experimental data were available.

As part of this investigation, the C~{\sc ii} lines from several observational line lists found in
the literature, such as that of Zhang \etal\ \cite{ZhangLLPB2005} for the planetary nebula
NGC~7027, were analyzed using our theoretical line list and all correctly-identified C\II\
recombination lines in these observational lists were identified in our theoretical list apart from
very few exceptions which are outside our wavelength range. The analysis also produced an electron
temperature for the line-emitting regions of a number of astronomical objects in reasonably good
agreement with the values obtained by other researchers using different data and employing other
techniques. The details of these investigations will be the subject of a forthcoming paper.

\section{Comparison to Previous Work} \label{Comparison}

In this section we make a brief comparison of some of our results against a sample of similar
results obtained by other researchers previously. These include radiative transition probabilities,
given in Table (\ref{TableRTP}), autoionization probabilities, given in Table (\ref{TableDSBAI}),
and dielectronic recombination coefficients, given in Table (\ref{TableBadnell}). Transition
probabilities generally show good agreement between the various calculations for the strongest
electric dipole transitions. There are some significant differences for intercombination
transitions, indicative of the increased uncertainty of the results for these cases. There are also
significant differences between the present results and those of Nussbaumer \& Storey
\cite{NussbaumerS19812} for some of the allowed but two-electron transitions from the 2s$^2$3p
configuration, where we would expect our results to be superior, given the larger scattering
target.

In Table (\ref{TableDSBAI}) we compare our calculated autoionization probabilities with those of De
Marco \etal\ \cite{DemarcoSB1997}. They combined the \LS-coupled autoionization probabilities
calculated in the close-coupling approximation by Davey \etal\ \cite{DaveySK2000} with one- and
two-body fine structure interactions computed with SUPERSTRUCTURE \cite{EissnerJN1974} to obtain
autoionization probabilities for four states that give rise to spectral lines seen in carbon-rich
Wolf-Rayet stars. For the three larger probabilities, there is agreement within 25\%. For the
4f~\SLPJ2Ge{9/2} state there is a factor of two difference but this state has a very small
autoionization probability corresponding to an energy width of only $4.2\times 10^{-8}$ Rydberg.

In Table (\ref{TableBadnell}) we compare our results with effective recombination coefficients from
Table~(1) of Badnell \cite{Badnell1988}. His results are tabulated between terms, having been
summed over the $J$ of the upper and lower terms of the transition, so results for individual lines
cannot be compared. Badnell only tabulates results for one transition in which the upper state is
allowed to autoionize in \LS-coupling (2s2p($^3$P$^{\rm o})$3d~$^2$F$^{\rm o}$) and here the
agreement is excellent as one might expect. On the other hand, the two levels of 2s2p($^3$P$^{\rm
o}$)3d~$^2$D$^{\rm o}$ have very small autoionization widths and our coefficients are factor of
~$5\times$ smaller than Badnell's for transitions from this term. We note that the fine structure
splitting of this term is well represented in our calculation as is its separation from neighboring
states with large autoionization widths, giving us confidence in our result.

\section{Results} \label{Results}
In this section, we present a sample of the data produced during this investigation. In Table
(\ref{BTable}) the theoretical results for the energies of the bound states of C$^+$ below the
C$^{2+}$ \SLPJ1Se0\ threshold are given alongside the available experimental data from the NIST
database \cite{NIST2010}. Similarly, Table (\ref{RTableKQ}) presents the energy and autoionization
width data for the resonances as obtained by the \Km-matrix and \QB\ methods. In these tables, a
negative fine structure splitting indicates that the theoretical levels are in reverse order
compared to their experimental counterparts. It is noteworthy that due to limited precision of
figures in these tables, some data may appear inconsistent, e.g. a zero fine structure splitting
from two levels with different energies. Full-precision data in electronic format are available
from the Centre de Donn\'{e}es astronomiques de Strasbourg (CDS) database.

Regarding the bound states, all levels with effective quantum number between 0.1-13 for the outer
electron and $0\le l \le 5$ (142 states) were found.  The 8 uppermost bound states in Table
(\ref{BTable}), i.e. the levels of 2s2p(\SLP3Po)3d~\SLP4Fo and \SLP4Do, have quantum numbers higher
than 13 and hence are out of range of the \Rm-matrix approximation validity; therefore only
experimental data are included for these states. Concerning the resonances, we searched for all
states with $n<5$ where $n$ is the principal quantum number of the active electron. 61 levels were
found by the \Km-matrix method and 55 by the \QB\ method.

Tables (\ref{fValues261o}-\ref{fValues265o}) present a sample of the $\OS$-values of the free-bound
transitions for some bound symmetries as obtained by integrating \pcs\ over photon energy where
these data are obtained from stage STGBF of the \Rm-matrix code.  The columns in these tables stand
for the bound states identified by their indices as given in Table (\ref{BTable}) while the rows
stand for the resonances represented by their indices as given in Table (\ref{RTableKQ}).
An entry of `0' in the $\OS$-value tables indicates that no peak was observed in the \pcs\ data.

Finally, Table (\ref{ListSample}) contains a sample of the effective recombination coefficients for
transitions extracted from our list in a wavelength range where several lines have been observed in
the spectra of planetary nebulae \cite{ZhangLLPB2005}.

\section{Conclusions} \label{Conclusions}

In this study, a list of line effective recombination coefficients is generated for the atomic ion
C$^+$ using the \Rm-matrix, \AS\ and Emissivity codes. These lines are produced by dielectronic
capture and subsequent radiative decays of the low-lying autoionizing states above the threshold of
C$^{2+}$ \SLP1Se with a principal quantum number $n<5$ for the captured electron. The line list
contains 6187 optically-allowed transitions which include many C~{\sc ii} lines observed in
astronomical spectra, notably the lines of C~{\sc ii} in the observational list of Zhang \etal\
\cite{ZhangLLPB2005} for the spectrum of planetary nebula NGC~7027. Beside the effective
recombination coefficients, the list also include detailed data of level energies for bound and
resonance states, and oscillator strengths.

The theoretical results for energy and fine structure splitting agree very well with the available
experimental data for both resonances and bound states.  The complete data set of our line list can
be obtained from the Centre de Donn\'{e}es astronomiques de Strasbourg (CDS) database.

In the course of this investigation, a method for finding and analyzing resonances was developed as
an alternative to the \QB\ \cite{QuigleyB1996, QuigleyBP1998} and Time-Delay \cite{StibbeT1998}
methods. In this one-channel case, the method offers a superior alternative in terms of numerical
stability, computational viability and comprehensiveness in the search for the low-lying
autoionizing states.

\clearpage
\section*{Figures}

\clearpage

\begin{table} [!b]
\caption{The seven orbitals used to construct the C$^{2+}$ target and their STO parameters. The bar
marks the pseudo orbitals.} \label{SevenorbitalsSTO} \centering
\begin{tabular}{cccc}
\hline
   Orbital &   $\STOCi$ &   $\STOPi$ &  $\STOxii$ \\
\hline
        1s &   21.28251 &          1 &    5.13180 \\
           &    6.37632 &          1 &    8.51900 \\
           &    0.08158 &          2 &    2.01880 \\
           &   -2.61339 &          2 &    4.73790 \\
           & -0.00733  \VS &          2 &    1.57130 \\
        2s &   -5.39193 &          1 &    5.13180 \\
           &   -1.49036 &          1 &    8.51900 \\
           &    5.57151 &          2 &    2.01880 \\
           &    -5.25090 &          2 &    4.73790 \\
           & 0.94247  \VS &          2 &    1.57130 \\
   $\po3s$ &    5.69321 &          1 &    1.75917 \\
           &  -19.54864 &          2 &    1.75917 \\
           & 10.39428  \VS &          3 &    1.75917 \\
        2p &    1.01509 &          2 &    1.47510 \\
           &    3.80119 &          2 &    3.19410 \\
           &    2.75006 &          2 &    1.83070 \\
           & 0.89571  \VS &          2 &    9.48450 \\
  $\po3p$ &   14.41203 &          2 &    1.98138 \\
           & -10.88586  \VS &          3 &    1.96954 \\
  $\po3d$ & 5.84915  \VS &          3 &    2.11997 \\
  $\po4f$ &    9.69136 &          4 &    2.69086 \\
\hline
\hspace{3cm} & \hspace{3cm} & \hspace{3cm} & \hspace{3cm} \\

\end{tabular}
\end{table}


\clearpage

\begin{table} [!h]
\caption{Orbital scaling parameters ($\lambda$'s) for \AS\ input. The rows stand for the principal
quantum number $n$, while the columns stand for the orbital angular momentum quantum number $l$.}
\label{lambdaTable} \vspace{0.2cm}

\centering
\begin{tabular}{|c|c|c|c|c|c|c|}
 \hline
 & s & p & d & f & g & h \\
 \hline
 1 & 1.43240 &  &  &  &  & \\
 \hline
 2 & 1.43380 & 1.39690 &  &  &  & \\
 \hline
 3 & 1.25760 & 1.20290 & 1.35930 &  &  & \\
 \hline
 4 & 1.25830 & 1.19950 & 1.35610 & 1.41460 &  & \\
 \hline
 5 & 1.26080 & 1.20020 & 1.35770 & 1.41420 & 1.32960 & \\
 \hline
 6 & 1.26370 & 1.20250 & 1.36210 & 1.41520 & 1.41420 & 2.34460 \\
 \hline
 7 & 1.26790 &  &  &  &  & \\
 \hline
\end{tabular}
\end{table}


\clearpage

\begin{table} [!h]
{\tiny

\vspace{-1.5cm}

\caption{A sample of radiative transition probabilities in s$^{-1}$ as obtained from this work
compared to corresponding values found in the literature. \label{TableRTP}}

\begin{center} \vspace{-0.2cm}

\begin{tabular}{|c|c|c|c|c|c|c|c|c|c|c|}
    \hline
    Upper & Lower & $\WL_{vac}$ & SS & NS & LDHK & HST & GKMKW & FKWP & DT & DSB \\
    \hline
    2s2p$^2$ \SLPJ4Pe{1/2} & 2s$^2$2p \SLPJ2Po{1/2} & 2325.40 & 52.8 & 55.3 & 74.4 &  &  &  & 42.5 &  \\
    \hline
    2s2p$^2$ \SLPJ4Pe{3/2} & 2s$^2$2p \SLPJ2Po{1/2} & 2324.21 & 1.79 & 1.71 & 1.70 &  &  &  & 1.01 &  \\
    \hline
    2s2p$^2$ \SLPJ4Pe{1/2} & 2s$^2$2p \SLPJ2Po{3/2} & 2328.84 & 60.6 & 65.5 & 77.8 &  &  &  & 40.2 &  \\
    \hline
    2s2p$^2$ \SLPJ4Pe{3/2} & 2s$^2$2p \SLPJ2Po{3/2} & 2327.64 & 9.34 & 5.24 & 12.4 &  &  &  & 8.11 &  \\
    \hline
    2s2p$^2$ \SLPJ4Pe{5/2} & 2s$^2$2p \SLPJ2Po{3/2} & 2326.11 & 36.7 & 43.2 & 53.9 &  &  & 51.2 & 34.4 &  \\
    \hline
    2s2p$^2$ \SLPJ2De{3/2} & 2s$^2$2p \SLPJ2Po{1/2} & 1334.53 & 2.40e8 & 2.42e8 & 2.38e8 &  &  &  & 2.41e8 &  \\
    \hline
    2s2p$^2$ \SLPJ2De{5/2} & 2s$^2$2p \SLPJ2Po{3/2} & 1335.71 & 2.87e8 & 2.88e8 & 2.84e8 &  &  &  & 2.89e8 &  \\
    \hline
    2s2p$^2$ \SLPJ2De{3/2} & 2s$^2$2p \SLPJ2Po{3/2} & 1335.66 & 4.75e7 & 4.78e7 & 4.70e7 &  &  &  & 4.79e7 &  \\
    \hline
    2s2p$^2$ \SLPJ2Se{1/2} & 2s$^2$2p \SLPJ2Po{1/2} & 1036.34 & 7.75e8 & 7.74e8 &  &  &  &  & 7.99e8 &  \\
    \hline
    2s2p$^2$ \SLPJ2Se{1/2} & 2s$^2$2p \SLPJ2Po{3/2} & 1037.02 & 1.53e9 & 1.53e9 &  &  &  &  & 1.59e9 &  \\
    \hline
    2s2p$^2$ \SLPJ2Pe{1/2} & 2s$^2$2p \SLPJ2Po{1/2} & 903.96 & 2.76e9 & 2.74e9  &  &  &  &  & 2.63e9 &  \\
    \hline
    2s2p$^2$ \SLPJ2Pe{3/2} & 2s$^2$2p \SLPJ2Po{1/2} & 903.62 & 6.92e8 & 6.86e8 &  &  &  &  & 6.58e8 &  \\
    \hline
    2s2p$^2$ \SLPJ2Pe{1/2} & 2s$^2$2p \SLPJ2Po{3/2} & 904.48 & 1.39e9 & 1.38e9 &  &  &  &  & 1.33e9 &  \\
    \hline
    2s2p$^2$ \SLPJ2Pe{3/2} & 2s$^2$2p \SLPJ2Po{3/2} & 904.14 & 3.46e9 & 3.43e9 &  &  &  &  & 3.30e9 &  \\
    \hline
    2s$^2$3s \SLPJ2Se{1/2} & 2s$^2$2p \SLPJ2Po{1/2} & 858.09 & 1.26e8 & 1.31e8 &  &  &  &  & 3.69e7 &  \\
    \hline
    2s$^2$3s \SLPJ2Se{1/2} & 2s$^2$2p \SLPJ2Po{3/2} & 858.56 & 2.46e8 & 2.58e8 &  &  &  &  & 1.11e8 &  \\
    \hline
    2s$^2$3p \SLPJ2Po{1/2} & 2s2p$^2$ \SLPJ4Pe{1/2} & 1127.13 & 36.3 & 55.5 &  &  &  &  &  &  \\
    \hline
    2s$^2$3p \SLPJ2Po{3/2} & 2s2p$^2$ \SLPJ4Pe{1/2} & 1126.99 & 58.28 & 97.8 &  &  &  &  &  &  \\
    \hline
    2s$^2$3p \SLPJ2Po{1/2} & 2s2p$^2$ \SLPJ4Pe{3/2} & 1127.41 & 21.8 & 20.3 &  &  &  &  &  &  \\
    \hline
    2s$^2$3p \SLPJ2Po{3/2} & 2s2p$^2$ \SLPJ4Pe{3/2} & 1127.27 & 14.0 & 22.3 &  &  &  &  &  &  \\
    \hline
    2s$^2$3p \SLPJ2Po{3/2} & 2s2p$^2$ \SLPJ4Pe{5/2} & 1127.63 & 1.86e2 & 2.60e2 &  &  &  &  &  &  \\
    \hline
    2p$^3$ \SLPJ4So{3/2} & 2s2p$^2$ \SLPJ4Pe{1/2} & 1009.86 & 5.77e8 & 5.82e8 &  &  &  &  & 5.76e8 &  \\
    \hline
    2p$^3$ \SLPJ4So{3/2} & 2s2p$^2$ \SLPJ4Pe{3/2} & 1010.08 & 1.15e9 & 1.16e9 &  &  &  &  & 1.15e9 &  \\
    \hline
    2p$^3$ \SLPJ4So{3/2} & 2s2p$^2$ \SLPJ4Pe{5/2} & 1010.37 & 1.73e9 & 1.74e9 &  &  &  &  & 1.73e9 &  \\
    \hline
    2s$^2$3p \SLPJ2Po{3/2} & 2s2p$^2$ \SLPJ2De{5/2} & 1760.40 & 3.77e7 & 3.75e7 &  & 4.3e7 &  &  &  &  \\
    \hline
    2s$^2$3p \SLPJ2Po{1/2} & 2s2p$^2$ \SLPJ2De{3/2} & 1760.82 & 4.19e7 & 4.16e7 &  & 5.2e7 &  &  &  &  \\
    \hline
    2s$^2$3p \SLPJ2Po{3/2} & 2s2p$^2$ \SLPJ2De{3/2} & 1760.47 & 4.18e6 & 4.16e6 &  & 4e6 &  &  &  &  \\
    \hline
    2p$^3$ \SLPJ4So{3/2} & 2s2p$^2$ \SLPJ2De{5/2} & 1490.38 & 1.03e2 & 45.0 &  &  &  &  &  &  \\
    \hline
    2p$^3$ \SLPJ4So{3/2} & 2s2p$^2$ \SLPJ2De{3/2} & 1490.44 & 10.5 & 11.9 &  &  &  &  &  &  \\
    \hline
    2s$^2$3p \SLPJ2Po{1/2} & 2s2p$^2$ \SLPJ2Se{1/2} & 2838.44 & 3.61e7 & 3.63e7 &  & 2.2e7 &  &  &  &  \\
    \hline
    2s$^2$3p \SLPJ2Po{3/2} & 2s2p$^2$ \SLPJ2Se{1/2} & 2837.54 & 3.61e7 & 3.64e7 &  & 2.2e7 &  &  &  &  \\
    \hline
    2p$^3$ \SLPJ4So{3/2} & 2s2p$^2$ \SLPJ2Se{1/2} & 2196.19 & 1.62 & 5.09 &  &  &  &  &  &  \\
    \hline
    2s$^2$3p \SLPJ2Po{1/2} & 2s2p$^2$ \SLPJ2Pe{1/2} & 4739.29 & 5.85e4 & 7.70e4 &  & $<$3.0e5 &  &  &  &  \\
    \hline
    2s$^2$3p \SLPJ2Po{3/2} & 2s2p$^2$ \SLPJ2Pe{1/2} & 4736.79 & 7.98e3 & 1.19e4 &  & $<$5e5 &  &  &  &  \\
    \hline
    2s$^2$3p \SLPJ2Po{1/2} & 2s2p$^2$ \SLPJ2Pe{3/2} & 4748.61 & 2.35e4 & 3.23e4 &  & $<$2.3e5 &  &  &  &  \\
    \hline
    2s$^2$3p \SLPJ2Po{3/2} & 2s2p$^2$ \SLPJ2Pe{3/2} & 4746.09 & 6.08e4 & 8.28e4 &  & $<$1.3e5 &  &  &  &  \\
    \hline
    2p$^3$ \SLPJ4So{3/2} & 2s2p$^2$ \SLPJ2Pe{1/2} & 3184.42 & 16.4 & 14.9 &  &  &  &  &  &  \\
    \hline
    2p$^3$ \SLPJ4So{3/2} & 2s2p$^2$ \SLPJ2Pe{3/2} & 3188.62 & 80.0 & 76.9 &  &  &  &  &  &  \\
    \hline
    2s$^2$3p \SLPJ2Po{1/2} & 2s$^2$3s \SLPJ2Se{1/2} & 6584.70 & 3.53e7 & 3.68e7 &  & 2.9e7 & 3.29e7 &  &  &  \\
    \hline
    2s$^2$3p \SLPJ2Po{3/2} & 2s$^2$3s \SLPJ2Se{1/2} & 6579.87 & 3.54e7 & 3.69e7 &  & 3.4e7 & 3.33e7 &  &  &  \\
    \hline
    2p$^3$ \SLPJ4So{3/2} & 2s$^2$3s \SLPJ2Se{1/2} & 3923.19 & 6.35 & 10.2 &  &  &  &  &  &  \\
    \hline
    2s$^2$3d \SLPJ2De{3/2} & 2s$^2$3p \SLPJ2Po{1/2} & 7233.33 & 3.46e7 &  &  &  & 4.22e7 &  &  &  \\
    \hline
    2s$^2$3d \SLPJ2De{3/2} & 2s$^2$3p \SLPJ2Po{3/2} & 7239.16 & 6.91e6 &  &  &  & 8.49e6 &  &  &  \\
    \hline
    2s$^2$3d \SLPJ2De{5/2} & 2s$^2$3p \SLPJ2Po{3/2} & 7238.41 & 4.15e7 &  &  &  & 5.10e7 &  &  &  \\
    \hline
    2s$^2$3d \SLPJ2De{3/2} & 2s$^2$2p \SLPJ2Po{1/2} & 687.05 & 2.40e9 &  &  &  &  &  & 2.24e9 &  \\
    \hline
    2s$^2$3d \SLPJ2De{3/2} & 2s$^2$2p \SLPJ2Po{3/2} & 687.35 & 4.81e8 &  &  &  &  &  & 4.50e8 &  \\
    \hline
    2s$^2$3d \SLPJ2De{5/2} & 2s$^2$2p \SLPJ2Po{3/2} & 687.35 & 2.88e9 &  &  &  &  &  & 2.70e9 &  \\
    \hline
    2p$^3$ \SLPJ2Do{3/2} & 2s2p$^2$ \SLPJ2De{3/2} & 1323.91 & 4.33e8 &  &  &  &  &  & 4.53e8 &  \\
    \hline
    2p$^3$ \SLPJ2Do{5/2} & 2s2p$^2$ \SLPJ2De{3/2} & 1324.00 & 3.23e7 &  &  &  &  &  & 3.38e7 &  \\
    \hline
    2p$^3$ \SLPJ2Do{3/2} & 2s2p$^2$ \SLPJ2De{5/2} & 1323.86 & 4.91e7 &  &  &  &  &  & 5.10e7 &  \\
    \hline
    2p$^3$ \SLPJ2Do{5/2} & 2s2p$^2$ \SLPJ2De{5/2} & 1323.95 & 4.51e8 &  &  &  &  &  & 4.71e8 &  \\
    \hline
    2p$^3$ \SLPJ2Do{3/2} & 2s2p$^2$ \SLPJ2Pe{1/2} & 2509.88 & 4.53e7 &  &  &  &  &  & 5.38e7 &  \\
    \hline
    2p$^3$ \SLPJ2Do{3/2} & 2s2p$^2$ \SLPJ2Pe{3/2} & 2512.49 & 8.91e6 &  &  &  &  &  & 1.06e7 &  \\
    \hline
    2p$^3$ \SLPJ2Do{5/2} & 2s2p$^2$ \SLPJ2Pe{3/2} & 2512.81 & 5.40e7 &  &  &  &  &  & 6.43e7 &  \\
    \hline
    2s2p(\SLP3Po)4f \SLPJ2De{5/2} & 2s2p(\SLP3Po)3d \SLPJ2Po{3/2} & 5115.07 & 1.18e8 &  &  &  &  &  &  & 6.94e7 \\
    \hline
    2s2p(\SLP3Po)4f \SLPJ2Ge{9/2} & 2s2p(\SLP3Po)3d \SLPJ2Fo{7/2} & 4620.54 & 2.29e8 &  &  &  &  &  &  & 1.84e8 \\
    \hline
    2s2p(\SLP3Po)3d \SLPJ2Po{3/2} & 2s2p(\SLP3Po)3p \SLPJ2Pe{3/2} & 4966.12 & 3.14e7 &  &  &  &  &  &  & 2.89e7 \\
    \hline
    2s2p(\SLP3Po)3d \SLPJ2Fo{7/2} & 2s2p(\SLP3Po)3p \SLPJ2De{5/2} & 8796.49 & 2.03e7 &  &  &  &  &  &  & 1.99e7 \\
    \hline
\end{tabular}

\end{center}
\vspace{0.0cm} {SS = current work, NS = Nussbaumer and Storey \cite{NussbaumerS19812}; LDHK =
Lennon \etal\ \cite{LennonDHK1985}; HST = Huber \etal\ \cite{HuberST1984}; GKMKW = Glenzer \etal\
\cite{GlenzerKMKW1994}; FKWP = Fang \etal\ \cite{FangKWP1993}; DT = Dankwort and Trefftz
\cite{DankwortT1978}, DSB = De Marco \etal\ \cite{DemarcoSB1997}.} }
\end{table}


\clearpage

\begin{table} [!h]
\caption{Autoionization probabilities in s$^{-1}$ of four resonance states as obtained in the
current work (SS) compared to those obtained by De Marco \etal\ \cite{DemarcoSB1997} (DSB).
\label{TableDSBAI}}

\begin{center}
\begin{tabular}{|c|c|c|}
    \hline
    State & SS & DSB \\
    \hline
    2s2p(\SLP3Po)4f \SLPJ2De{5/2} & 9.860e10 & 8.263e10 \\
    \hline
    2s2p(\SLP3Po)4f \SLPJ2Ge{9/2} & 8.601e8 & 1.623e9 \\
    \hline
    2s2p(\SLP3Po)3d \SLPJ2Po{3/2} & 2.194e11 & 1.901e11 \\
    \hline
    2s2p(\SLP3Po)3d \SLPJ2Fo{7/2} & 1.196e12 & 1.488e12 \\
    \hline
\end{tabular}
\end{center}
\end{table}


\clearpage

{\scriptsize

\begin{longtable}{@{\extracolsep\fill}llllllllllllr@{}}
\caption{Effective dielectronic recombination rate coefficients in cm$^3$.s$^{-1}$ of a number of
transitions for the given logarithmic temperatures. The first row for each transition corresponds
to Badnell \cite{Badnell1988}, while the second row is obtained from the current work. The
superscripts denote powers of 10.
\label{TableBadnell}} \vspace{-0.7cm}\\

           &            &            &            &            &            & {\bf log10(T)} &            &            &            &            &            \\

{\bf Upper} & {\bf Lower} &  {\bf 3.0} &  {\bf 3.2} &  {\bf 3.4} &  {\bf 3.6} &  {\bf 3.8} &  {\bf 4.0} &  {\bf 4.2} &  {\bf 4.4} &  {\bf 4.6} &  {\bf 4.8} \\

2s2p(\SLP3Po)3d \SLP4Fo & 2s2p(\SLP3Po)3p \SLP4De &   7.27\EE{-24} &   5.73\EE{-20} &   1.56\EE{-17} &   6.17\EE{-16} &   6.47\EE{-15} &   2.61\EE{-14} &   5.52\EE{-14} &   7.43\EE{-14} &   7.25\EE{-14} &   5.65\EE{-14} \\

           &            &   6.74\EE{-24} &   5.17\EE{-20} &   1.29\EE{-17} &   4.33\EE{-16} &   3.90\EE{-15} &   1.33\EE{-14} &   2.31\EE{-14} &   2.57\EE{-14} &   2.13\EE{-14} &   1.47\EE{-14} \\

2s2p(\SLP3Po)3p \SLP4De & 2s2p(\SLP3Po)3s \SLP4Po &   1.90\EE{-16} &   3.21\EE{-16} &   8.53\EE{-16} &   4.95\EE{-15} &   1.93\EE{-14} &   4.90\EE{-14} &   8.51\EE{-14} &   1.05\EE{-13} &   9.91\EE{-14} &   7.57\EE{-14} \\

           &            &   1.06\EE{-16} &   1.90\EE{-16} &   7.71\EE{-16} &   5.02\EE{-15} &   1.74\EE{-14} &   3.58\EE{-14} &   4.85\EE{-14} &   4.76\EE{-14} &   3.70\EE{-14} &   2.46\EE{-14} \\

2s2p(\SLP3Po)3d \SLP2Do & 2s2p(\SLP3Po)3p \SLP2Pe &   2.49\EE{-14} &   3.19\EE{-14} &   2.90\EE{-14} &   2.11\EE{-14} &   1.35\EE{-14} &   8.05\EE{-15} &   4.80\EE{-15} &   2.94\EE{-15} &   1.81\EE{-15} &   1.10\EE{-15} \\

           &            &   4.67\EE{-15} &   5.99\EE{-15} &   5.44\EE{-15} &   3.97\EE{-15} &   2.58\EE{-15} &   1.68\EE{-15} &   1.17\EE{-15} &   8.31\EE{-16} &   5.54\EE{-16} &   3.41\EE{-16} \\

2s2p(\SLP3Po)3p \SLP4Pe & 2s2p(\SLP3Po)3s \SLP4Po &   3.08\EE{-16} &   5.09\EE{-16} &   9.86\EE{-16} &   4.02\EE{-15} &   1.08\EE{-14} &   1.73\EE{-14} &   2.04\EE{-14} &   1.97\EE{-14} &   1.61\EE{-14} &   1.14\EE{-14} \\

           &            &   1.49\EE{-16} &   2.52\EE{-16} &   7.24\EE{-16} &   3.84\EE{-15} &   1.05\EE{-14} &   1.58\EE{-14} &   1.60\EE{-14} &   1.27\EE{-14} &   8.52\EE{-15} &   5.14\EE{-15} \\

2s2p(\SLP3Po)4f \SLP2Fe & 2s2p(\SLP3Po)3d \SLP2Do &   3.00\EE{-27} &   6.39\EE{-22} &   1.14\EE{-18} &   9.93\EE{-17} &   1.29\EE{-15} &   5.09\EE{-15} &   9.43\EE{-15} &   1.09\EE{-14} &   9.26\EE{-15} &   6.51\EE{-15} \\

           &            &   1.64\EE{-27} &   3.53\EE{-22} &   6.33\EE{-19} &   5.54\EE{-17} &   7.22\EE{-16} &   2.82\EE{-15} &   5.18\EE{-15} &   5.88\EE{-15} &   4.94\EE{-15} &   3.43\EE{-15} \\

2s2p(\SLP3Po)4f \SLP4De & 2s2p(\SLP3Po)3d \SLP4Po &   2.56\EE{-27} &   7.68\EE{-22} &   1.70\EE{-18} &   1.70\EE{-16} &   2.40\EE{-15} &   9.95\EE{-15} &   1.90\EE{-14} &   2.21\EE{-14} &   1.89\EE{-14} &   1.33\EE{-14} \\

           &            &   2.54\EE{-27} &   7.63\EE{-22} &   1.69\EE{-18} &   1.69\EE{-16} &   2.39\EE{-15} &   9.86\EE{-15} &   1.87\EE{-14} &   2.17\EE{-14} &   1.85\EE{-14} &   1.29\EE{-14} \\

2s2p(\SLP3Po)4s \SLP4Po & 2s2p(\SLP3Po)3p \SLP4Pe &   2.59\EE{-20} &   1.23\EE{-17} &   4.65\EE{-16} &   3.58\EE{-15} &   1.01\EE{-14} &   1.51\EE{-14} &   1.51\EE{-14} &   1.18\EE{-14} &   7.80\EE{-15} &   4.67\EE{-15} \\

           &            &   2.63\EE{-20} &   1.25\EE{-17} &   4.72\EE{-16} &   3.63\EE{-15} &   1.02\EE{-14} &   1.53\EE{-14} &   1.53\EE{-14} &   1.19\EE{-14} &   7.83\EE{-15} &   4.68\EE{-15} \\

2s2p(\SLP3Po)4f \SLP4Fe & 2s2p(\SLP3Po)3d \SLP4Do &   5.39\EE{-27} &   1.16\EE{-21} &   2.07\EE{-18} &   1.81\EE{-16} &   2.37\EE{-15} &   9.41\EE{-15} &   1.77\EE{-14} &   2.07\EE{-14} &   1.79\EE{-14} &   1.27\EE{-14} \\

           &            &   3.48\EE{-27} &   7.51\EE{-22} &   1.35\EE{-18} &   1.19\EE{-16} &   1.55\EE{-15} &   6.07\EE{-15} &   1.11\EE{-14} &   1.26\EE{-14} &   1.06\EE{-14} &   7.37\EE{-15} \\

2s2p(\SLP3Po)4f \SLP4Ge & 2s2p(\SLP3Po)3d \SLP4Fo &   5.42\EE{-27} &   1.50\EE{-21} &   3.15\EE{-18} &   3.05\EE{-16} &   4.24\EE{-15} &   1.74\EE{-14} &   3.30\EE{-14} &   3.86\EE{-14} &   3.32\EE{-14} &   2.34\EE{-14} \\

           &            &   2.98\EE{-27} &   8.25\EE{-22} &   1.73\EE{-18} &   1.68\EE{-16} &   2.33\EE{-15} &   9.49\EE{-15} &   1.78\EE{-14} &   2.06\EE{-14} &   1.75\EE{-14} &   1.22\EE{-14} \\

2s2p(\SLP3Po)4s \SLP4Po & 2s2p(\SLP3Po)3p \SLP4De &   2.86\EE{-20} &   1.36\EE{-17} &   5.14\EE{-16} &   3.95\EE{-15} &   1.11\EE{-14} &   1.67\EE{-14} &   1.67\EE{-14} &   1.30\EE{-14} &   8.61\EE{-15} &   5.15\EE{-15} \\

           &            &   3.15\EE{-20} &   1.49\EE{-17} &   5.65\EE{-16} &   4.35\EE{-15} &   1.22\EE{-14} &   1.83\EE{-14} &   1.83\EE{-14} &   1.42\EE{-14} &   9.37\EE{-15} &   5.59\EE{-15} \\

2s2p(\SLP3Po)4d \SLP4Fo & 2s2p(\SLP3Po)3p \SLP4De &   1.77\EE{-26} &   1.71\EE{-21} &   1.85\EE{-18} &   1.18\EE{-16} &   1.30\EE{-15} &   4.85\EE{-15} &   9.35\EE{-15} &   1.18\EE{-14} &   1.10\EE{-14} &   8.31\EE{-15} \\

           &            &   1.75\EE{-26} &   1.69\EE{-21} &   1.83\EE{-18} &   1.17\EE{-16} &   1.24\EE{-15} &   4.28\EE{-15} &   7.25\EE{-15} &   7.83\EE{-15} &   6.37\EE{-15} &   4.34\EE{-15} \\

2s2p$^2$ \SLP2De & 2s$^2$2p \SLP2Po &   2.45\EE{-12} &   5.89\EE{-12} &   8.34\EE{-12} &   8.17\EE{-12} &   6.33\EE{-12} &   4.36\EE{-12} &   3.01\EE{-12} &   2.21\EE{-12} &   1.66\EE{-12} &   1.19\EE{-12} \\

           &            &   1.85\EE{-12} &   5.12\EE{-12} &   7.66\EE{-12} &   7.67\EE{-12} &   6.00\EE{-12} &   4.09\EE{-12} &   2.61\EE{-12} &   1.60\EE{-12} &   9.38\EE{-13} &   5.28\EE{-13} \\

2s2p(\SLP3Po)3d \SLP2Do & 2s2p$^2$ \SLP2Pe &   2.60\EE{-13} &   3.33\EE{-13} &   3.02\EE{-13} &   2.20\EE{-13} &   1.41\EE{-13} &   8.41\EE{-14} &   5.01\EE{-14} &   3.07\EE{-14} &   1.89\EE{-14} &   1.15\EE{-14} \\

           &            &   4.50\EE{-14} &   5.78\EE{-14} &   5.24\EE{-14} &   3.83\EE{-14} &   2.49\EE{-14} &   1.62\EE{-14} &   1.13\EE{-14} &   8.01\EE{-15} &   5.34\EE{-15} &   3.29\EE{-15} \\

2s2p(\SLP3Po)4d \SLP2Do & 2s2p$^2$ \SLP2Pe &   6.56\EE{-27} &   1.10\EE{-21} &   1.70\EE{-18} &   1.35\EE{-16} &   1.66\EE{-15} &   6.33\EE{-15} &   1.16\EE{-14} &   1.33\EE{-14} &   1.13\EE{-14} &   8.02\EE{-15} \\

           &            &   1.98\EE{-27} &   3.29\EE{-22} &   5.03\EE{-19} &   3.97\EE{-17} &   4.85\EE{-16} &   1.82\EE{-15} &   3.26\EE{-15} &   3.64\EE{-15} &   3.03\EE{-15} &   2.09\EE{-15} \\

2s2p$^2$ \SLP2Pe & 2s$^2$2p \SLP2Po &   2.68\EE{-13} &   3.99\EE{-13} &   5.05\EE{-13} &   5.45\EE{-13} &   4.87\EE{-13} &   3.95\EE{-13} &   3.37\EE{-13} &   3.09\EE{-13} &   2.75\EE{-13} &   2.21\EE{-13} \\

           &            &   5.11\EE{-14} &   1.11\EE{-13} &   2.21\EE{-13} &   3.09\EE{-13} &   3.18\EE{-13} &   2.78\EE{-13} &   2.24\EE{-13} &   1.66\EE{-13} &   1.11\EE{-13} &   6.82\EE{-14} \\

2s2p(\SLP3Po)3d \SLP2Do & 2s2p$^2$ \SLP2De &   7.38\EE{-13} &   9.47\EE{-13} &   8.59\EE{-13} &   6.26\EE{-13} &   3.99\EE{-13} &   2.39\EE{-13} &   1.42\EE{-13} &   8.71\EE{-14} &   5.38\EE{-14} &   3.25\EE{-14} \\

           &            &   1.36\EE{-13} &   1.75\EE{-13} &   1.59\EE{-13} &   1.16\EE{-13} &   7.53\EE{-14} &   4.90\EE{-14} &   3.43\EE{-14} &   2.42\EE{-14} &   1.61\EE{-14} &   9.95\EE{-15} \\

2s2p(\SLP3Po)3s \SLP4Po & 2s2p$^2$ \SLP4Pe &   1.36\EE{-15} &   2.22\EE{-15} &   3.36\EE{-15} &   1.07\EE{-14} &   3.30\EE{-14} &   7.13\EE{-14} &   1.12\EE{-13} &   1.32\EE{-13} &   1.21\EE{-13} &   9.09\EE{-14} \\

           &            &   7.17\EE{-16} &   1.19\EE{-15} &   2.35\EE{-15} &   1.02\EE{-14} &   3.12\EE{-14} &   5.78\EE{-14} &   7.22\EE{-14} &   6.71\EE{-14} &   5.05\EE{-14} &   3.29\EE{-14} \\

2s2p(\SLP3Po)3d \SLP2Fo & 2s2p$^2$ \SLP2De &   1.70\EE{-12} &   4.92\EE{-12} &   7.44\EE{-12} &   7.49\EE{-12} &   5.83\EE{-12} &   3.86\EE{-12} &   2.30\EE{-12} &   1.29\EE{-12} &   6.93\EE{-13} &   3.63\EE{-13} \\

           &            &   1.71\EE{-12} &   4.94\EE{-12} &   7.48\EE{-12} &   7.53\EE{-12} &   5.86\EE{-12} &   3.88\EE{-12} &   2.32\EE{-12} &   1.30\EE{-12} &   6.97\EE{-13} &   3.65\EE{-13} \\

2s2p(\SLP3Po)4d \SLP2Do & 2s2p$^2$ \SLP2De &   2.30\EE{-26} &   3.86\EE{-21} &   5.94\EE{-18} &   4.72\EE{-16} &   5.81\EE{-15} &   2.22\EE{-14} &   4.05\EE{-14} &   4.65\EE{-14} &   3.97\EE{-14} &   2.81\EE{-14} \\

           &            &   9.59\EE{-27} &   1.59\EE{-21} &   2.43\EE{-18} &   1.92\EE{-16} &   2.35\EE{-15} &   8.82\EE{-15} &   1.58\EE{-14} &   1.76\EE{-14} &   1.47\EE{-14} &   1.01\EE{-14} \\

2s2p(\SLP3Po)3d \SLP4Do & 2s2p$^2$ \SLP4Pe &   2.49\EE{-24} &   2.83\EE{-20} &   1.05\EE{-17} &   4.56\EE{-16} &   4.63\EE{-15} &   1.78\EE{-14} &   3.62\EE{-14} &   4.76\EE{-14} &   4.58\EE{-14} &   3.53\EE{-14} \\

           &            &   2.34\EE{-24} &   2.63\EE{-20} &   9.38\EE{-18} &   3.84\EE{-16} &   3.62\EE{-15} &   1.23\EE{-14} &   2.11\EE{-14} &   2.32\EE{-14} &   1.91\EE{-14} &   1.31\EE{-14} \\

2s2p(\SLP3Po)3d \SLP4Po & 2s2p$^2$ \SLP4Pe &   6.71\EE{-14} &   1.08\EE{-13} &   1.13\EE{-13} &   9.07\EE{-14} &   6.36\EE{-14} &   4.86\EE{-14} &   4.61\EE{-14} &   4.51\EE{-14} &   3.87\EE{-14} &   2.84\EE{-14} \\

           &            &   3.41\EE{-14} &   5.51\EE{-14} &   5.78\EE{-14} &   4.63\EE{-14} &   3.38\EE{-14} &   2.97\EE{-14} &   3.10\EE{-14} &   2.93\EE{-14} &   2.30\EE{-14} &   1.55\EE{-14} \\

2s2p(\SLP3Po)4s \SLP4Po & 2s2p$^2$ \SLP4Pe &   9.12\EE{-20} &   4.32\EE{-17} &   1.64\EE{-15} &   1.26\EE{-14} &   3.55\EE{-14} &   5.31\EE{-14} &   5.33\EE{-14} &   4.14\EE{-14} &   2.75\EE{-14} &   1.64\EE{-14} \\

           &            &   1.13\EE{-19} &   5.37\EE{-17} &   2.03\EE{-15} &   1.56\EE{-14} &   4.40\EE{-14} &   6.58\EE{-14} &   6.59\EE{-14} &   5.12\EE{-14} &   3.38\EE{-14} &   2.02\EE{-14} \\

2s2p(\SLP3Po)4d \SLP4Do & 2s2p$^2$ \SLP4Pe &   4.87\EE{-26} &   6.33\EE{-21} &   8.27\EE{-18} &   5.94\EE{-16} &   6.90\EE{-15} &   2.57\EE{-14} &   4.72\EE{-14} &   5.54\EE{-14} &   4.85\EE{-14} &   3.49\EE{-14} \\

           &            &   1.73\EE{-26} &   2.26\EE{-21} &   2.95\EE{-18} &   2.12\EE{-16} &   2.43\EE{-15} &   8.79\EE{-15} &   1.53\EE{-14} &   1.69\EE{-14} &   1.39\EE{-14} &   9.52\EE{-15} \\

2s2p(\SLP3Po)4d \SLP4Po & 2s2p$^2$ \SLP4Pe &   5.09\EE{-27} &   9.49\EE{-22} &   1.56\EE{-18} &   1.30\EE{-16} &   1.68\EE{-15} &   6.90\EE{-15} &   1.39\EE{-14} &   1.78\EE{-14} &   1.67\EE{-14} &   1.27\EE{-14} \\

           &            &   4.37\EE{-28} &   8.18\EE{-23} &   1.34\EE{-19} &   1.11\EE{-17} &   1.40\EE{-16} &   5.37\EE{-16} &   9.70\EE{-16} &   1.09\EE{-15} &   9.13\EE{-16} &   6.32\EE{-16} \\

2s2p(\SLP3Po)3p \SLP2Pe & 2s$^2$2p \SLP2Po &   1.91\EE{-14} &   2.70\EE{-14} &   3.11\EE{-14} &   3.17\EE{-14} &   2.97\EE{-14} &   2.65\EE{-14} &   2.26\EE{-14} &   1.81\EE{-14} &   1.34\EE{-14} &   9.18\EE{-15} \\

           &            &   4.59\EE{-15} &   9.59\EE{-15} &   1.86\EE{-14} &   2.67\EE{-14} &   2.99\EE{-14} &   2.83\EE{-14} &   2.31\EE{-14} &   1.66\EE{-14} &   1.06\EE{-14} &   6.26\EE{-15} \\

2s2p(\SLP3Po)4p \SLP2Pe & 2s$^2$2p \SLP2Po &   5.81\EE{-23} &   3.67\EE{-19} &   7.12\EE{-17} &   1.53\EE{-15} &   8.24\EE{-15} &   1.85\EE{-14} &   2.40\EE{-14} &   2.19\EE{-14} &   1.61\EE{-14} &   1.03\EE{-14} \\

           &            &   6.35\EE{-23} &   3.96\EE{-19} &   7.62\EE{-17} &   1.63\EE{-15} &   8.74\EE{-15} &   1.96\EE{-14} &   2.52\EE{-14} &   2.30\EE{-14} &   1.68\EE{-14} &   1.07\EE{-14} \\

\end{longtable}
}

\renewcommand{\baselinestretch}{1.0}

\newpage


\datatables

\setlength{\LTleft}{0pt}
\setlength{\LTright}{0pt}
\setlength{\tabcolsep}{0.5\tabcolsep}
\renewcommand{\arraystretch}{1.0}


\clearpage

{\footnotesize

\begin{longtable}{@{\extracolsep\fill}lllllllr@{}}
\caption{The available experimental data from NIST for the bound states of C$^+$ below the C$^{2+}$
\SLPJ1Se0\ threshold alongside the theoretical results from \Rm-matrix\ calculations.
\label{BTable}} \\
\hline\hline %
{In.$^a$} & {Config.$^b$} & {Level} & {NEEW$^c$} & {NEER$^d$} & {FSS$^e$} & {TER$^f$} & {FSSR$^g$} \\ %
\hline
\endfirsthead
\caption[]{continued.}\\
\hline\hline %
{In.} & {Config.} & {Level} & {NEEW} & {NEER} & {FSS} & {TER} & {FSSR} \\ %
\hline
\endhead
\hline
\endfoot

1   &   2s$^2$2p    &   \SLPJ2Po{1/2}   &   0   &   -1.792141   &       &   -1.792571   &       \\
2   &   2s$^2$2p    &    \SLPJ2Po{3/2}  &   63.42   &   -1.791563   &   63.4    &   -1.791994   &   63.3    \VS\\

3   &   2s2p$^2$    &    \SLPJ4Pe{1/2}  &   43003.3 &   -1.400266   &       &   -1.401292   &       \\
4   &   2s2p$^2$    &    \SLPJ4Pe{3/2}  &   43025.3 &   -1.400065   &       &   -1.401090   &       \\
5   &   2s2p$^2$    &    \SLPJ4Pe{5/2}  &   43053.6 &   -1.399807   &   50.3    &   -1.400755   &   59.0    \VS\\

6   &   2s2p$^2$    &    \SLPJ2De{5/2}  &   74930.1 &   -1.109327   &       &   -1.105247   &       \\
7   &   2s2p$^2$    &    \SLPJ2De{3/2}  &   74932.62    &   -1.109304   &   2.5 &   -1.105266   &   -2.1    \VS\\

8   &   2s2p$^2$    &    \SLPJ2Se{1/2}  &   96493.74    &   -0.912825   &       &   -0.899439   &       \VS\\

9   &   2s2p$^2$    &    \SLPJ2Pe{1/2}  &   110624.17   &   -0.784059   &       &  -0.772978   &       \\
10  &   2s2p$^2$    &    \SLPJ2Pe{3/2}  &   110665.56   &   -0.783682   &   41.4    &   -0.772570   &   44.8    \VS\\

11  &   2s$^2$3s    &    \SLPJ2Se{1/2}  &   116537.65   &   -0.730171   &       &   -0.727698   &       \VS\\

12  &   2s$^2$3p    &    \SLPJ2Po{1/2}  &   131724.37   &   -0.591780   &       &   -0.592081   &       \\
13  &   2s$^2$3p    &    \SLPJ2Po{3/2}  &   131735.52   &   -0.591678   &   11.1    &   -0.591980   &   11.1    \VS\\

14  &   2p$^3$  &    \SLPJ4So{3/2}  &   142027.1    &   -0.497894   &       &   -0.487491   &       \VS\\

15  &   2s$^2$3d    &    \SLPJ2De{3/2}  &   145549.27   &   -0.465798   &       &   -0.466024   &       \\
16  &   2s$^2$3d    &    \SLPJ2De{5/2}  &   145550.7    &   -0.465785   &   1.4 &   -0.466005   &   2.1 \VS\\

17  &   2p$^3$  &    \SLPJ2Do{5/2}  &   150461.58   &   -0.421034   &       &   -0.412359   &       \\
18  &   2p$^3$  &    \SLPJ2Do{3/2}  &   150466.69   &   -0.420987   &   5.1 &   -0.412388   &   -3.1    \VS\\

19  &   2s$^2$4s    &    \SLPJ2Se{1/2}  &   157234.07   &   -0.359318   &       &   -0.358955   &       \VS\\

20  &   2s$^2$4p    &    \SLPJ2Po{1/2}  &   162517.89   &   -0.311169   &       &   -0.310784   &       \\
21  &   2s$^2$4p    &    \SLPJ2Po{3/2}  &   162524.57   &   -0.311108   &   6.7 &   -0.310723   &   6.6 \VS\\

22  &   2s2p(\SLP3Po)3s     &    \SLPJ4Po{1/2}  &   166967.13   &   -0.270624   &       &   -0.268893   &       \\
23  &   2s2p(\SLP3Po)3s     &    \SLPJ4Po{3/2}  &   166990.73   &   -0.270409   &       &   -0.268645   &       \\
24  &   2s2p(\SLP3Po)3s     &    \SLPJ4Po{5/2}  &   167035.71   &   -0.269999   &   68.6    &   -0.268229   &   72.9    \VS\\

25  &   2s$^2$4d    &    \SLPJ2De{3/2}  &   168123.74   &   -0.260084   &       &   -0.260084   &       \\
26  &   2s$^2$4d    &    \SLPJ2De{5/2}  &   168124.45   &   -0.260078   &   0.7 &   -0.260074   &   1.1 \VS\\

27  &   2p$^3$  &    \SLPJ2Po{1/2}  &   168729.53   &   -0.254564   &       &   -0.245329   &       \\
28  &   2p$^3$  &    \SLPJ2Po{3/2}  &   168748.3    &   -0.254393   &   18.8    &   -0.245083   &   27.1    \VS\\

29  &   2s$^2$4f    &    \SLPJ2Fo{5/2}  &   168978.34   &   -0.252297   &       &   -0.252162   &       \\
30  &   2s$^2$4f    &    \SLPJ2Fo{7/2}  &   168978.34   &   -0.252297   &   0.0 &   -0.252160   &   0.2 \VS\\

31  &   2s$^2$5s    &    \SLPJ2Se{1/2}  &   173347.84   &   -0.212479   &       &   -0.212278   &       \VS\\

32  &   2s$^2$5p    &    \SLPJ2Po{1/2}  &   175287.39   &   -0.194804   &       &   -0.190099   &       \\
33  &   2s$^2$5p    &    \SLPJ2Po{3/2}  &   175294.75   &   -0.194737   &   7.4 &   -0.190072   &   3.0 \VS\\

34  &   2s2p(\SLP3Po)3s     &    \SLPJ2Po{1/2}  &   177774.59   &   -0.172139   &       &   -0.165330   &       \\
35  &   2s2p(\SLP3Po)3s     &    \SLPJ2Po{3/2}  &   177793.54   &   -0.171967   &   19.0    &   -0.165143   &   20.5    \VS\\

36  &   2s$^2$5d    &    \SLPJ2De{3/2}  &   178495.11   &   -0.165573   &       &   -0.165606   &       \\
37  &   2s$^2$5d    &    \SLPJ2De{5/2}  &   178495.71   &   -0.165568   &   0.6 &   -0.165599   &   0.8 \VS\\

38  &   2s$^2$5f    &    \SLPJ2Fo{5/2}  &   178955.94   &   -0.161374   &       &   -0.161296   &       \\
39  &   2s$^2$5f    &    \SLPJ2Fo{7/2}  &   178955.94   &   -0.161374   &   0.0 &   -0.161295   &   0.1 \VS\\

40  &   2s$^2$5g    &    \SLPJ2Ge{7/2}  &   179073.05   &   -0.160307   &       &   -0.160184   &       \\
41  &   2s$^2$5g    &    \SLPJ2Ge{9/2}  &   179073.05   &   -0.160307   &   0.0 &   -0.160184   &   0.0 \VS\\

42  &   2s$^2$6s    &    \SLPJ2Se{1/2}  &   181264.24   &   -0.140339   &       &   -0.140231   &       \VS\\

43  &   2s2p(\SLP3Po)3p     &    \SLPJ4De{1/2}  &   181696.66   &   -0.136399   &       &   -0.136848   &       \\
44  &   2s2p(\SLP3Po)3p     &    \SLPJ4De{3/2}  &   181711.03   &   -0.136268   &       &   -0.136699   &       \\
45  &   2s2p(\SLP3Po)3p     &    \SLPJ4De{5/2}  &   181736.05   &   -0.136040   &       &   -0.136454   &       \\
46  &   2s2p(\SLP3Po)3p     &    \SLPJ4De{7/2}  &   181772.41   &   -0.135709   &   75.8    &   -0.136115   &   80.4    \VS\\

47  &   2s2p(\SLP3Po)3p     &    \SLPJ2Pe{1/2}  &   182023.86   &   -0.133417   &       &   -0.133095   &       \\
48  &   2s2p(\SLP3Po)3p     &    \SLPJ2Pe{3/2}  &   182043.41   &   -0.133239   &   19.6    &   -0.132896   &   21.8    \VS\\

49  &   2s$^2$6p    &    \SLPJ2Po{1/2}  &   182993.23   &   -0.124584   &       &   -0.124376   &       \\
50  &   2s$^2$6p    &    \SLPJ2Po{3/2}  &   182993.66   &   -0.124580   &   0.4 &   -0.124351   &   2.7 \VS\\

51  &   2s$^2$6d    &    \SLPJ2De{3/2}  &   184074.59   &   -0.114730   &       &   -0.114739   &       \\
52  &   2s$^2$6d    &    \SLPJ2De{5/2}  &   184075.28   &   -0.114723   &   0.7 &   -0.114729   &   1.0 \VS\\

53  &   2s$^2$6f    &    \SLPJ2Fo{5/2}  &   184376.06   &   -0.111982   &       &   -0.111924   &       \\
54  &   2s$^2$6f    &    \SLPJ2Fo{7/2}  &   184376.06   &   -0.111982   &   0.0 &   -0.111924   &   0.1 \VS\\

55  &   2s$^2$6g    &    \SLPJ2Ge{7/2}  &   184449.27   &   -0.111315   &       &   -0.111264   &       \\
56  &   2s$^2$6g    &    \SLPJ2Ge{9/2}  &   184449.27   &   -0.111315   &   0.0 &   -0.111264   &   0.0 \VS\\

57  &   2s$^2$6h    &    \SLPJ2Ho{9/2}  &   184466.5    &   -0.111158   &       &   -0.111122   &       \\
58  &   2s$^2$6h    &    \SLPJ2Ho{11/2} &   184466.5    &   -0.111158   &   0.0 &   -0.111122   &   0.0 \VS\\

59  &   2s2p(\SLP3Po)3p     &    \SLPJ4Se{3/2}  &   184690.98   &   -0.109113   &       &   -0.108410   &       \VS\\

60  &   2s$^2$7s    &    \SLPJ2Se{1/2}  &   185732.93   &   -0.099618   &       &   -0.099537   &       \VS\\

61  &   2s2p(\SLP3Po)3p     &    \SLPJ4Pe{1/2}  &   186427.35   &   -0.093290   &       &   -0.092097   &       \\
62  &   2s2p(\SLP3Po)3p     &    \SLPJ4Pe{3/2}  &   186443.69   &   -0.093141   &       &   -0.091950   &       \\
63  &   2s2p(\SLP3Po)3p     &    \SLPJ4Pe{5/2}  &   186466.02   &   -0.092937   &   38.7    &   -0.091717   &   41.8    \VS\\

64  &   2s$^2$7p    &    \SLPJ2Po{1/2}  &   186745.9    &   -0.090387   &       &   -0.090313   &       \\
65  &   2s$^2$7p    &    \SLPJ2Po{3/2}  &   186746.3    &   -0.090383   &   0.4 &   -0.090302   &   1.1 \VS\\

66  &   2s$^2$7d    &    \SLPJ2De{3/2}  &   187353  &   -0.084854   &       &   -0.084804   &       \\
67  &   2s$^2$7d    &    \SLPJ2De{5/2}  &   187353  &   -0.084854   &   0.0 &   -0.084764   &   4.4 \VS\\

68  &   2s$^2$7f    &    \SLPJ2Fo{5/2}  &   187641.6    &   -0.082225   &       &   -0.082171   &       \\
69  &   2s$^2$7f    &    \SLPJ2Fo{7/2}  &   187641.6    &   -0.082225   &   0.0 &   -0.082171   &   0.0 \VS\\

70  &   2s$^2$7g    &    \SLPJ2Ge{7/2}  &   187691.4    &   -0.081771   &       &   -0.081744   &       \\
71  &   2s$^2$7g    &    \SLPJ2Ge{9/2}  &   187691.4    &   -0.081771   &   0.0 &   -0.081744   &   0.0 \VS\\

72  &   2s$^2$7h    &    \SLPJ2Ho{9/2}  &   187701  &   -0.081683   &       &   -0.081645   &       \\
73  &   2s$^2$7h    &    \SLPJ2Ho{11/2} &   187701  &   -0.081683   &   0.0 &   -0.081645   &   0.0 \VS\\

74  &   2s$^2$8s    &    \SLPJ2Se{1/2}  &   --- &   --- &       &   -0.074324   &       \VS\\

75  &   2s2p(\SLP3Po)3p     &    \SLPJ2De{3/2}  &   188581.25   &   -0.073662   &       &   -0.071901   &       \\
76  &   2s2p(\SLP3Po)3p     &    \SLPJ2De{5/2}  &   188615.07   &   -0.073354   &   33.8    &   -0.071582   &   34.9    \VS\\

77  &   2s$^2$8p    &    \SLPJ2Po{1/2}  &   --- &   --- &       &   -0.068349   &       \\
78  &   2s$^2$8p    &    \SLPJ2Po{3/2}  &   --- &   --- &       &   -0.068343   &   0.7 \VS\\

79  &   2s$^2$8f    &    \SLPJ2Fo{5/2}  &   --- &   --- &       &   -0.062873   &       \\
80  &   2s$^2$8f    &    \SLPJ2Fo{7/2}  &   --- &   --- &       &   -0.062872   &   0.0 \VS\\

81  &   2s$^2$8g    &    \SLPJ2Ge{7/2}  &   189794.2    &   -0.062609   &       &   -0.062580   &       \\
82  &   2s$^2$8g    &    \SLPJ2Ge{9/2}  &   189794.2    &   -0.062609   &   0.0 &   -0.062580   &   0.0 \VS\\

83  &   2s$^2$8h    &    \SLPJ2Ho{11/2} &   --- &   --- &       &   -0.062511   &       \\
84  &   2s$^2$8h    &    \SLPJ2Ho{9/2}  &   --- &   --- &       &   -0.062511   &   0.0 \VS\\

85  &   2s$^2$8d    &   \SLPJ2De{3/2}   &   --- &   --- &       &   -0.062612   &       \\
86  &   2s$^2$8d    &   \SLPJ2De{5/2}   &   --- &   --- &       &   -0.062564   &   5.3 \VS\\

87  &   2s$^2$9s    &   \SLPJ2Se{1/2}   &   --- &   --- &       &   -0.057638   &       \VS\\

88  &   2s$^2$9p    &   \SLPJ2Po{1/2}   &   --- &   --- &       &   -0.053490   &       \\
89  &   2s$^2$9p    &   \SLPJ2Po{3/2}   &   --- &   --- &       &   -0.053486   &   0.4 \VS\\

90  &   2s$^2$9d    &   \SLPJ2De{3/2}   &   --- &   --- &       &   -0.049942   &       \\
91  &   2s$^2$9d    &   \SLPJ2De{5/2}   &   --- &   --- &       &   -0.049935   &   0.8 \VS\\

92  &   2s$^2$9f    &   \SLPJ2Fo{5/2}   &   --- &   --- &       &   -0.049651   &       \\
93  &   2s$^2$9f    &   \SLPJ2Fo{7/2}   &   --- &   --- &       &   -0.049650   &   0.0 \VS\\

94  &   2s$^2$9g    &   \SLPJ2Ge{7/2}   &   --- &   --- &       &   -0.049441   &       \\
95  &   2s$^2$9g    &   \SLPJ2Ge{9/2}   &   --- &   --- &       &   -0.049441   &   0.0 \VS\\

96  &   2s$^2$9h    &   \SLPJ2Ho{9/2}   &   --- &   --- &       &   -0.049392   &       \\
97  &   2s$^2$9h    &   \SLPJ2Ho{11/2}  &   --- &   --- &       &   -0.049392   &   0.0 \VS\\

98  &   2s$^2$10s   &   \SLPJ2Se{1/2}   &   --- &   --- &       &   -0.046050   &       \VS\\

99  &   2s$^2$10p   &   \SLPJ2Po{1/2}   &   --- &   --- &       &   -0.042990   &       \\
100 &   2s$^2$10p   &   \SLPJ2Po{3/2}   &   --- &   --- &       &  -0.042987   &   0.3 \VS\\

101 &   2s$^2$10d   &   \SLPJ2De{3/2}   &   --- &   --- &       &   -0.040487   &       \\
102 &   2s$^2$10d   &   \SLPJ2De{5/2}   &   --- &   --- &       &   -0.040484   &   0.3 \VS\\

103 &   2s$^2$10f   &   \SLPJ2Fo{5/2}   &   --- &   --- &       &   -0.040198   &       \\
104 &   2s$^2$10f   &   \SLPJ2Fo{7/2}   &   --- &   --- &       &   -0.040198   &   0.0 \VS\\

105 &   2s$^2$10g   &   \SLPJ2Ge{7/2}   &   --- &   --- &       &   -0.040044   &       \\
106 &   2s$^2$10g   &   \SLPJ2Ge{9/2}   &   --- &   --- &       &   -0.040044   &   0.0 \VS\\

107 &   2s$^2$10h   &   \SLPJ2Ho{9/2}   &   --- &   --- &       &   -0.040007   &       \\
108 &   2s$^2$10h   &   \SLPJ2Ho{11/2}  &   --- &   --- &       &   -0.040007   &   0.0 \VS\\

109 &   2s$^2$11s   &   \SLPJ2Se{1/2}   &   --- &   --- &       &   -0.037746   &       \VS\\

110 &   2s$^2$11p   &   \SLPJ2Po{1/2}   &   --- &   --- &       &   -0.035300   &       \\
111 &   2s$^2$11p   &   \SLPJ2Po{3/2}   &   --- &   --- &       &   -0.035298   &   0.2 \VS\\

112 &   2s$^2$11d   &   \SLPJ2De{3/2}   &   --- &   --- &       &   -0.033449   &       \\
113 &   2s$^2$11d   &   \SLPJ2De{5/2}   &   --- &   --- &       &   -0.033448   &   0.2 \VS\\

114 &   2s$^2$11f   &   \SLPJ2Fo{5/2}   &   --- &   --- &       &   -0.033209   &       \\
115 &   2s$^2$11f   &   \SLPJ2Fo{7/2}   &   --- &   --- &       &   -0.033209   &   0.0 \VS\\

116 &   2s$^2$11g   &   \SLPJ2Ge{7/2}   &   --- &   --- &       &   -0.033092   &       \\
117 &   2s$^2$11g   &   \SLPJ2Ge{9/2}   &   --- &   --- &       &   -0.033092   &   0.0 \VS\\

118 &   2s$^2$11h   &   \SLPJ2Ho{9/2}   &   --- &   --- &       &   -0.033064   &       \\
119 &   2s$^2$11h   &   \SLPJ2Ho{11/2}  &   --- &   --- &       &   -0.033064   &   0.0 \VS\\

120 &   2s$^2$12s   &   \SLPJ2Se{1/2}   &   --- &   --- &       &   -0.031953   &       \VS\\

121 &   2s$^2$12p   &   \SLPJ2Po{1/2}   &   --- &   --- &       &   -0.029501   &       \\
122 &   2s$^2$12p   &   \SLPJ2Po{3/2}   &   --- &   --- &       &   -0.029499   &   0.2 \VS\\

123 &   2s2p(\SLP3Po)3p &   \SLPJ2Se{1/2}   &   --- &   --- &       &   -0.028972   &       \VS\\

124 &   2s$^2$12d   &   \SLPJ2De{3/2}   &   --- &   --- &       &   -0.028090   &       \\
125 &   2s$^2$12d   &   \SLPJ2De{5/2}   &   --- &   --- &       &   -0.028089   &   0.1 \VS\\

126 &   2s$^2$12f   &   \SLPJ2Fo{5/2}   &   --- &   --- &       &   -0.027895   &       \\
127 &   2s$^2$12f   &   \SLPJ2Fo{7/2}   &   --- &   --- &       &   -0.027895   &   0.0 \VS\\

128 &   2s$^2$12g   &   \SLPJ2Ge{7/2}   &   --- &   --- &       &   -0.027804   &       \\
129 &   2s$^2$12g   &   \SLPJ2Ge{9/2}   &   --- &   --- &       &   -0.027804   &   0.0 \VS\\

130 &   2s$^2$12h   &   \SLPJ2Ho{9/2}   &   --- &   --- &       &   -0.027783   &       \\
131 &   2s$^2$12h   &   \SLPJ2Ho{11/2}  &   --- &   --- &       &   -0.027783   &   0.0 \VS\\

132 &   2s$^2$13s   &   \SLPJ2Se{1/2}   &   --- &   --- &       &   -0.025758   &       \VS\\

133 &   2s$^2$13p   &   \SLPJ2Po{1/2}   &   --- &   --- &       &   -0.025021   &       \\
134 &   2s$^2$13p   &   \SLPJ2Po{3/2}   &   --- &   --- &       &   -0.025020   &   0.1 \VS\\

135 &   2s$^2$13d   &   \SLPJ2De{3/2}   &   --- &   --- &       &   -0.023920   &       \\
136 &   2s$^2$13d   &   \SLPJ2De{5/2}   &   --- &   --- &       &   -0.023919   &   0.1 \VS\\

137 &   2s$^2$13f   &   \SLPJ2Fo{5/2}   &   --- &   --- &       &   -0.023762   &       \\
138 &   2s$^2$13f   &   \SLPJ2Fo{7/2}   &   --- &   --- &       &   -0.023762   &   0.0 \VS\\

139 &   2s$^2$13g   &   \SLPJ2Ge{7/2}   &   --- &   --- &       &   -0.023690   &       \\
140 &   2s$^2$13g   &   \SLPJ2Ge{9/2}   &   --- &   --- &       &   -0.023690   &   0.0 \VS\\

141 &   2s$^2$13h   &   \SLPJ2Ho{9/2}   &   --- &   --- &       &   -0.023673   &       \\
142 &   2s$^2$13h   &   \SLPJ2Ho{11/2}  &   --- &   --- &       &   -0.023673   &   0.0 \VS\\

143 &   2s2p(\SLP3Po)3d     &    \SLPJ4Fo{3/2}  &   195752.58   &   -0.008312   &       &       &       \\
144 &   2s2p(\SLP3Po)3d     &    \SLPJ4Fo{5/2}  &   195765.85   &   -0.008191   &       &       &       \\
145 &   2s2p(\SLP3Po)3d     &    \SLPJ4Fo{7/2}  &   195785.74   &   -0.008010   &       &       &       \\
146 &   2s2p(\SLP3Po)3d     &    \SLPJ4Fo{9/2}  &   195813.66   &   -0.007755   &   61.1    &       &       \VS\\

147 &   2s2p(\SLP3Po)3d     &    \SLPJ4Do{1/2}  &   196557.87   &   -0.000974   &       &       &       \\
148 &   2s2p(\SLP3Po)3d     &    \SLPJ4Do{3/2}  &   196563.41   &   -0.000923   &       &       &       \\
149 &   2s2p(\SLP3Po)3d     &    \SLPJ4Do{5/2}  &   196571.82   &   -0.000846   &       &       &       \\
150 &   2s2p(\SLP3Po)3d     &    \SLPJ4Do{7/2}  &   196581.96   &   -0.000754   &   24.1    &       &       \VS\\

\end{longtable}
\begin{list}{}{}
\item[$^{a}$] Index.
\item[$^{b}$] Configuration. The 1s$^{2}$ core is suppressed from all configurations.
\item[$^{c}$] NIST Experimental Energy in Wavenumbers (cm$^{-1}$) relative to the ground state.
\item[$^{d}$] NIST Experimental Energy in Rydberg relative to the C$^{2+}$ \SLPJ1Se0\ limit.
\item[$^{e}$] Fine Structure Splitting from experimental values in cm$^{-1}$. %
\item[$^{f}$] Theoretical Energy in Rydberg from \Rm-matrix calculations relative to the C$^{2+}$ \SLPJ1Se0\ limit. %
\item[$^{g}$] Fine Structure Splitting from \Rm-matrix in cm$^{-1}$. The minus sign indicates that the theoretical levels are in reverse order compared to the experimental. %
\end{list}

}


\clearpage

{\small

\begin{longtable}{@{\extracolsep\fill}llllllllllll@{}}
\caption{The available experimental data from NIST for the resonance states of C$^+$ above the
C$^{2+}$ \SLPJ1Se0\ threshold alongside the theoretical results as obtained by \Km-matrix and \QB\
methods.\label{RTableKQ}}\\
\hline\hline %
{In.$^a$} & {Config.$^b$} & {Level} & {NEEW$^c$} & {NEER$^d$} & {FSS$^e$} & {TERK$^f$} & {FSSK$^g$} & {FWHMK$^h$} & {TERQ$^i$} & {FSSQ$^j$} & {FWHMQ$^k$} \\ %
\hline
\endfirsthead
\caption[]{continued.}\\
\hline\hline %
{In.} & {Config.} & {Level} & {NEEW} & {NEER} & {FSS} & {TERK} & {FSSK} & {FWHMK} & {TERQ} & {FSSQ} & {FWHMQ} \\ %
\hline
\endhead
\hline
\endfoot

1   &   2s2p(\SLP3Po)3d     &    \SLPJ2Do{3/2}  &   198425.43   &   0.016045    &       &   0.017012    &       &   6.00E-10    &   0.017012    &       &   6.00E-10    \\
2   &   2s2p(\SLP3Po)3d     &    \SLPJ2Do{5/2}  &   198436.31   &   0.016144    &   10.9    &   0.017124    &   12.3    &   2.61E-09    &   0.017124    &   12.3    &  2.61E-09    \VS\\

3   &   2s2p(\SLP3Po)3d     &    \SLPJ4Po{5/2}  &   198844  &   0.019859    &       &   0.020655    &       &   2.17E-11    &    ---    &       &    ---    \\
4   &   2s2p(\SLP3Po)3d     &    \SLPJ4Po{3/2}  &   198865.25   &   0.020053    &       &   0.020849    &       &   1.26E-09    &   0.020849    &       &   1.26E-09    \\
5   &   2s2p(\SLP3Po)3d     &    \SLPJ4Po{1/2}  &   198879.01   &   0.020178    &   35.0    &   0.020968    &   34.3    &   5.10E-10    &   0.020968    &       &   5.10E-10    \VS\\

6   &   2s2p(\SLP3Po)3d     &    \SLPJ2Fo{5/2}  &   199941.41   &   0.029860    &       &   0.031693    &       &   5.61E-05    &   0.031689    &       &   5.61E-05    \\
7   &   2s2p(\SLP3Po)3d     &    \SLPJ2Fo{7/2}  &   199983.24   &   0.030241    &   41.8    &   0.032109    &   45.6    &   5.79E-05    &   0.032106    &   45.8    &   5.79E-05    \VS\\

8   &   2s2p(\SLP3Po)3d     &    \SLPJ2Po{3/2}  &   202179.85   &   0.050258    &       &   0.053787    &       &   1.06E-05    &   0.053787    &       &   1.06E-05    \\
9   &   2s2p(\SLP3Po)3d     &    \SLPJ2Po{1/2}  &   202204.52   &   0.050483    &   24.7    &   0.054019    &   25.5    &   1.12E-05    &   0.054019    &   25.5    &   1.12E-05    \VS\\

10  &   2s2p(\SLP3Po)4s     &    \SLPJ4Po{1/2}  &   209552.36   &   0.117441    &       &   0.118682    &       &   2.55E-07    &   0.118682    &       &   2.55E-07    \\
11  &   2s2p(\SLP3Po)4s     &    \SLPJ4Po{3/2}  &   209576.46   &   0.117661    &       &   0.118938    &       &   6.25E-07    &   0.118938    &       &   6.25E-07    \\
12  &   2s2p(\SLP3Po)4s     &    \SLPJ4Po{5/2}  &   209622.32   &   0.118079    &   70.0    &   0.119304    &   68.3    &   $<$1.0E-16  &    ---    &       &    ---    \VS\\

13  &   2s2p(\SLP3Po)4s     &    \SLPJ2Po{1/2}  &    ---    &    ---    &       &   0.137883    &       &  6.13E-03    &   0.137866    &       &   6.02E-03    \\
14  &   2s2p(\SLP3Po)4s     &    \SLPJ2Po{3/2}  &    ---    &    ---    &       &   0.138385    &   55.0    &   6.15E-03    &   0.138365    &   54.8    &   6.03E-03    \VS\\

15  &   2s2p(\SLP3Po)4p     &    \SLPJ2Pe{1/2}  &   214404.33   &   0.161655    &       &   0.162743    &       &   2.12E-08    &   0.162743    &       &   2.12E-08    \\
16  &   2s2p(\SLP3Po)4p     &    \SLPJ2Pe{3/2}  &   214429.95   &   0.161889    &   25.6    &   0.162996    &   27.7    &   9.89E-08    &   0.162996    &   27.8    &   9.89E-08    \VS\\

17  &   2s2p(\SLP3Po)4p     &    \SLPJ4De{1/2}  &   214759.91   &   0.164896    &       &   0.165629    &       &   5.30E-11    &   0.165629    &       &   5.31E-11    \\
18  &   2s2p(\SLP3Po)4p     &    \SLPJ4De{3/2}  &   214772.84   &   0.165014    &       &   0.165760    &       &   5.12E-08    &   0.165760    &       &   5.12E-08    \\
19  &   2s2p(\SLP3Po)4p     &    \SLPJ4De{5/2}  &   214795.27   &   0.165218    &       &   0.165984    &       &   6.95E-08    &   0.165984    &       &   6.95E-08    \\
20  &   2s2p(\SLP3Po)4p     &    \SLPJ4De{7/2}  &   214829.77   &   0.165532    &   69.9    &    ---    &       &    ---    &    ---    &       &   ---    \VS\\

21  &   2s2p(\SLP3Po)4p     &    \SLPJ4Se{3/2}  &   215767.77   &   0.174080    &       &  0.174625    &       &   6.59E-12    &    ---    &       &    ---    \VS\\

22  &   2s2p(\SLP3Po)4p     &    \SLPJ4Pe{1/2}  &   216362.84   &   0.179503    &       &   0.180780    &       &   7.26E-08    &   0.180780    &       &   7.26E-08    \\
23  &   2s2p(\SLP3Po)4p     &    \SLPJ4Pe{3/2}  &   216379.59   &   0.179655    &       &   0.180931    &       &   1.03E-07    &   0.180931    &       &   1.03E-07    \\
24  &   2s2p(\SLP3Po)4p     &    \SLPJ4Pe{5/2}  &   216400.57   &   0.179846    &   37.7    &   0.181145    &   40.0    &   6.05E-07    &   0.181145    &   40.0    &   6.05E-07    \VS\\

25  &   2s2p(\SLP3Po)4p     &    \SLPJ2De{3/2}  &    ---    &    ---    &       &   0.185420    &       &   1.19E-03    &   0.185422    &       &   1.18E-03    \\
26  &   2s2p(\SLP3Po)4p     &    \SLPJ2De{5/2}  &   216927  &   0.184644    &       &   0.185859    &   48.2    &   1.18E-03    &   0.185859    &   48.0    &   1.18E-03    \VS\\

27  &   2s2p(\SLP3Po)4p     &    \SLPJ2Se{1/2}  &    ---    &    ---    &       &   0.199765    &       &   4.28E-04    &   0.199765    &       &   4.28E-04    \VS\\

28  &   2s2p(\SLP1Po)3s     &    \SLPJ2Po{1/2}  &    ---    &    ---    &       &   0.205892    &       &   3.03E-03    &   0.205889    &       &   3.02E-03    \\
29  &   2s2p(\SLP1Po)3s     &    \SLPJ2Po{3/2}  &    ---    &    ---    &       &   0.205912    &   2.2 &   3.01E-03    &   0.205909    &   2.2 &   3.01E-03    \VS\\

30  &   2s2p(\SLP3Po)4d     &    \SLPJ4Fo{3/2}  &   219556.54   &   0.208606    &       &   0.208824    &       &   5.08E-11    &   0.208824    &       &   5.08E-11    \\
31  &   2s2p(\SLP3Po)4d     &    \SLPJ4Fo{5/2}  &   219570.15   &   0.208730    &       &   0.208968    &       &   4.23E-09    &   0.208968    &       &   4.23E-09    \\
32  &   2s2p(\SLP3Po)4d     &    \SLPJ4Fo{7/2}  &   219590.76   &   0.208918    &       &   0.209174    &       &   5.96E-09    &   0.209174    &       &   5.96E-09    \\
33  &   2s2p(\SLP3Po)4d     &    \SLPJ4Fo{9/2}  &   219619.88   &   0.209183    &   63.3    &    ---    &       &    ---    &    ---    &       &    ---    \VS\\

34  &   2s2p(\SLP3Po)4d     &    \SLPJ4Do{1/2}  &   220125.51   &   0.213791    &       &   0.214804    &       &   2.31E-09    &   0.214804    &       &   2.31E-09    \\
35  &   2s2p(\SLP3Po)4d     &    \SLPJ4Do{3/2}  &   220130.86   &   0.213839    &       &   0.214847    &       &   1.06E-09    &   0.214847    &       &   1.06E-09    \\
36  &   2s2p(\SLP3Po)4d     &    \SLPJ4Do{5/2}  &   220139.41   &   0.213917    &       &   0.214923    &       &   1.33E-09    &   0.214923    &       &   1.33E-09    \\
37  &   2s2p(\SLP3Po)4d     &    \SLPJ4Do{7/2}  &   220150.49   &   0.214018    &   25.0    &   0.215042    &   26.1    &   6.19E-09    &   0.215042    &   26.1    &   6.19E-09    \VS\\

38  &   2s2p(\SLP3Po)4d     &    \SLPJ2Do{3/2}  &   220601.53   &   0.218128    &       &   0.219325    &       &   1.99E-08    &   0.219325    &       &   1.99E-08    \\
39  &   2s2p(\SLP3Po)4d     &    \SLPJ2Do{5/2}  &   220614.51   &   0.218247    &   13.0    &   0.219441    &   12.7    &   5.68E-09    &   0.219441    &   12.7    &   5.68E-09    \VS\\

40  &   2s2p(\SLP3Po)4d     &    \SLPJ4Po{5/2}  &   220811.69   &   0.220044    &       &   0.220495    &       &   2.43E-10    &   0.220495    &       &   2.43E-10    \\
41  &   2s2p(\SLP3Po)4d     &    \SLPJ4Po{3/2}  &   220832.15   &   0.220230    &       &   0.220680    &       &   5.32E-10    &   0.220680    &       &   5.32E-10    \\
42  &   2s2p(\SLP3Po)4d     &    \SLPJ4Po{1/2}  &   220845.07   &   0.220348    &   33.4    &   0.220796    &   33.0    &   3.65E-10    &   0.220796    &   33.0    &   3.65E-10    \VS\\

43  &   2s2p(\SLP3Po)4f     &    \SLPJ2Fe{5/2}  &   221088.88   &   0.222570    &       &   0.223542    &       &   9.13E-09    &   0.223542    &       &   9.13E-09    \\
44  &   2s2p(\SLP3Po)4f     &    \SLPJ2Fe{7/2}  &   221097.92   &   0.222652    &   9.0 &   0.223626    &   9.2 &   1.98E-11    &    ---    &       &    ---    \VS\\

45  &   2s2p(\SLP3Po)4f     &    \SLPJ4Fe{3/2}  &   221093.95   &   0.222616    &       &   0.223626    &       &   5.96E-09    &   0.223626    &       &   5.96E-09    \\
46  &   2s2p(\SLP3Po)4f     &    \SLPJ4Fe{5/2}  &   221099.11   &   0.222663    &       &   0.223658    &       &   5.90E-09    &   0.223658    &       &   5.90E-09    \\
47  &   2s2p(\SLP3Po)4f     &    \SLPJ4Fe{7/2}  &   221105.73   &   0.222723    &       &   0.223725    &       &   5.43E-10    &   0.223725    &       &   5.43E-10    \\
48  &   2s2p(\SLP3Po)4f     &    \SLPJ4Fe{9/2}  &   221109.78   &   0.222760    &   15.8    &   0.223779    &   16.8    &   1.26E-09    &   0.223779    &   16.8    &   1.26E-09    \VS\\

49  &   2s2p(\SLP3Po)4d     &    \SLPJ2Fo{5/2}  &   221460.88   &   0.225959    &       &   0.227144    &       &   1.45E-05    &   0.227144    &       &   1.45E-05    \\
50  &   2s2p(\SLP3Po)4d     &    \SLPJ2Fo{7/2}  &   221503.33   &   0.226346    &   42.4    &   0.227556    &   45.1    &   1.51E-05    &   0.227556    &   45.1    &   1.51E-05    \VS\\

51  &   2s2p(\SLP3Po)4f     &    \SLPJ4Ge{5/2}  &   221544.81   &   0.226724    &       &   0.227648    &       &   5.45E-10    &   0.227648    &       &   5.46E-10    \\
52  &   2s2p(\SLP3Po)4f     &    \SLPJ4Ge{7/2}  &   221553.99   &   0.226808    &       &   0.227757    &       &   7.22E-09    &   0.227757    &       &   7.22E-09    \\
53  &   2s2p(\SLP3Po)4f     &    \SLPJ4Ge{9/2}  &   221575.61   &   0.227005    &       &   0.227966    &       &   6.53E-09    &   0.227966    &       &   6.54E-09    \\
54  &   2s2p(\SLP3Po)4f     &    \SLPJ4Ge{11/2} &   221604.9    &   0.227272    &   60.1    &    ---    &       &    ---    &    ---    &       &    ---    \VS\\

55  &   2s2p(\SLP3Po)4f     &    \SLPJ2Ge{7/2}  &   221587.12   &   0.227110    &       &   0.228164    &       &   2.58E-08    &   0.228164    &       &   2.58E-08    \\
56  &   2s2p(\SLP3Po)4f     &    \SLPJ2Ge{9/2}  &   221625.72   &   0.227462    &   38.6    &   0.228541    &   41.4    &   4.16E-08    &   0.228541    &   41.4    &   4.16E-08    \VS\\

57  &   2s2p(\SLP3Po)4f     &    \SLPJ4De{7/2}  &   221698.48   &   0.228125    &       &   0.229105    &       &   2.97E-12    &    ---    &       &   ---     \\
58  &   2s2p(\SLP3Po)4f     &    \SLPJ4De{5/2}  &   221707.71   &   0.228209    &       &   0.229229    &       &   2.39E-06    &   0.229229    &       &   2.39E-06    \\
59  &   2s2p(\SLP3Po)4f     &    \SLPJ4De{3/2}  &   221729.69   &   0.228409    &       &   0.229418    &       &   7.35E-07    &   0.229418    &       &   7.35E-07    \\
60  &   2s2p(\SLP3Po)4f     &    \SLPJ4De{1/2}  &   221742.79   &   0.228528    &   44.3    &   0.229529    &   46.5    &   $<$1.0E-16  &    ---    &       &    ---    \VS\\

61  &   2s2p(\SLP3Po)4f     &    \SLPJ2De{5/2}  &   221729.92   &   0.228411    &       &   0.229459    &       &   4.77E-06    &   0.229459    &       &   4.77E-06    \\
62  &   2s2p(\SLP3Po)4f     &    \SLPJ2De{3/2}  &   221752.26   &   0.228615    &   22.3    &   0.229695    &   25.9    &   6.58E-06    &   0.229696    &   25.9    &   6.58E-06    \VS\\

63  &   2s2p(\SLP3Po)4d     &    \SLPJ2Po{3/2}  &   222258.8    &   0.233231    &       &   0.234626    &       &   2.42E-05    &   0.234626    &       &   2.42E-05    \\
64  &   2s2p(\SLP3Po)4d     &    \SLPJ2Po{1/2}  &   222285.7    &   0.233476    &   26.9    &   0.234870    &   26.8    &   2.46E-05    &   0.234870    &   26.8    &   2.46E-05    \VS\\

\end{longtable}
\begin{list}{}{}
\item[$^{a}$] Index.
\item[$^{b}$] Configuration. The 1s$^{2}$ core is suppressed from all configurations.
\item[$^{c}$] NIST Experimental Energy in Wavenumbers (cm$^{-1}$) relative to the ground state.
\item[$^{d}$] NIST Experimental Energy in Rydberg relative to the C$^{2+}$ \SLPJ1Se0\ limit.
\item[$^{e}$] Fine Structure Splitting from experimental values in cm$^{-1}$. %
\item[$^{f}$] Theoretical Energy in Rydberg from \Km-matrix calculations relative to the C$^{2+}$ \SLPJ1Se0\ limit. %
\item[$^{g}$] Fine Structure Splitting from \Km-matrix in cm$^{-1}$. %
\item[$^{h}$] Full Width at Half Maximum from \Km-matrix in Rydberg. %
\item[$^{i}$] Theoretical Energy in Rydberg from \QB\ calculations relative to the C$^{2+}$ \SLPJ1Se0\ limit. %
\item[$^{j}$] Fine Structure Splitting from \QB\ in cm$^{-1}$. %
\item[$^{k}$] Full Width at Half Maximum from \QB\ in Rydberg. %
\end{list}

}


\clearpage

\begin{longtable}{l@{ }l@{ }l@{ }l@{ }l@{ }l@{ }l@{ }l@{ }l@{ }l@{ }l@{ }l@{ }l@{ }l@{ }l@{ }l}
\caption{Free-bound $\OS$-values for bound symmetry $J^{\Par} = 1/2^{\rm{o}}$ obtained by
integrating \pcs s from \Rm-matrix calculations. The superscript denotes the power of 10
by which the number is to be multiplied. \label{fValues261o}}\\
\hline\hline %
In.&                \cC{1}&             \cC{12}&            \cC{20}&            \cC{22}&            \cC{27}&            \cC{32}&            \cC{34}&            \cC{49}&            \cC{64}&            \cC{77}&            \cC{88}&            \cC{99}&            \cC{110}&           \cC{121}&           \cC{133}\\
\hline %
\endfirsthead %
\caption[]{continued.}\\ %
\hline\hline %
In.&                \cC{1}&             \cC{12}&            \cC{20}&            \cC{22}&            \cC{27}&            \cC{32}&            \cC{34}&            \cC{49}&            \cC{64}&            \cC{77}&            \cC{88}&            \cC{99}&            \cC{110}&           \cC{121}&           \cC{133}\\
\hline %
\endhead %
\hline %
\endfoot %
\vspace{-0.2cm} \\
15&                 6.12\EE{-3}&        1.57\EE{-3}&        5.60\EE{-4}&        4.58\EE{-5}&        1.92\EE{-2}&        1.01\EE{-3}&        2.89\EE{-2}&        3.93\EE{-3}&        1.55\EE{-3}&        8.99\EE{-4}&        5.98\EE{-4}&        4.26\EE{-4}&        3.17\EE{-4}&        2.44\EE{-4}&        1.92\EE{-4}\\
16&                 3.18\EE{-3}&        8.30\EE{-4}&        2.69\EE{-4}&        1.21\EE{-5}&        9.72\EE{-3}&        5.02\EE{-4}&        1.47\EE{-2}&        2.02\EE{-3}&        7.92\EE{-4}&        4.63\EE{-4}&        3.07\EE{-4}&        2.20\EE{-4}&        1.64\EE{-4}&        1.26\EE{-4}&        9.93\EE{-5}\\
17&                 1.70\EE{-5}&        4.69\EE{-6}&        1.59\EE{-6}&        1.34\EE{-2}&        5.91\EE{-5}&        2.92\EE{-6}&        8.14\EE{-5}&        1.09\EE{-5}&        4.27\EE{-6}&        2.47\EE{-6}&        1.64\EE{-6}&        1.16\EE{-6}&        8.63\EE{-7}&        6.60\EE{-7}&        5.15\EE{-7}\\
18&                 8.53\EE{-6}&        3.43\EE{-6}&        4.94\EE{-7}&        1.35\EE{-2}&        1.19\EE{-5}&        7.05\EE{-7}&        2.68\EE{-5}&        5.00\EE{-6}&        1.95\EE{-6}&        1.19\EE{-6}&        8.00\EE{-7}&        5.89\EE{-7}&        4.64\EE{-7}&        3.50\EE{-7}&        2.90\EE{-7}\\
21&                 4.43\EE{-7}&        1.39\EE{-7}&        5.34\EE{-8}&        1.78\EE{-3}&        1.67\EE{-6}&        7.21\EE{-8}&        2.04\EE{-6}&        2.71\EE{-7}&        1.04\EE{-7}&        5.92\EE{-8}&        3.88\EE{-8}&        2.71\EE{-8}&        1.99\EE{-8}&        1.51\EE{-8}&        1.18\EE{-8}\\
22&                 4.48\EE{-7}&        3.87\EE{-7}&        4.64\EE{-9}&        2.22\EE{-4}&        6.57\EE{-13}&       7.26\EE{-8}&        8.82\EE{-7}&        1.54\EE{-7}&        7.90\EE{-8}&        5.31\EE{-8}&        3.85\EE{-8}&        2.95\EE{-8}&        2.41\EE{-8}&        1.91\EE{-8}&        1.53\EE{-8}\\
23&                 2.44\EE{-6}&        1.65\EE{-6}&        6.16\EE{-7}&        1.01\EE{-3}&        1.65\EE{-7}&        5.61\EE{-8}&        7.62\EE{-7}&        5.76\EE{-7}&        3.75\EE{-7}&        2.74\EE{-7}&        2.08\EE{-7}&        1.61\EE{-7}&        1.27\EE{-7}&        1.02\EE{-7}&        8.32\EE{-8}\\
25&                 1.81\EE{-2}&        1.13\EE{-2}&        3.63\EE{-3}&        3.44\EE{-7}&        2.22\EE{-3}&        1.01\EE{-3}&        1.66\EE{-2}&        7.00\EE{-3}&        4.20\EE{-3}&        2.96\EE{-3}&        2.20\EE{-3}&        1.69\EE{-3}&        1.32\EE{-3}&        1.05\EE{-3}&        8.50\EE{-4}\\
27&                 1.81\EE{-3}&        2.76\EE{-3}&        7.42\EE{-4}&        3.83\EE{-8}&        1.55\EE{-3}&        4.26\EE{-4}&        2.16\EE{-3}&        4.12\EE{-4}&        2.69\EE{-4}&        1.98\EE{-4}&        1.52\EE{-4}&        1.20\EE{-4}&        9.55\EE{-5}&        7.73\EE{-5}&        6.32\EE{-5}\\
45&                 1.31\EE{-9}&        3.51\EE{-8}&        1.80\EE{-8}&        7.18\EE{-7}&        3.61\EE{-6}&        6.21\EE{-6}&        1.84\EE{-5}&        6.64\EE{-7}&        1.40\EE{-7}&        6.59\EE{-8}&        4.25\EE{-8}&        2.98\EE{-8}&        2.26\EE{-8}&        1.82\EE{-8}&        1.53\EE{-8}\\
59&                 8.42\EE{-7}&        3.47\EE{-6}&        1.84\EE{-6}&        6.24\EE{-4}&        4.20\EE{-4}&        6.82\EE{-4}&        1.99\EE{-3}&        6.74\EE{-5}&        1.46\EE{-5}&        6.42\EE{-6}&        3.64\EE{-6}&        2.34\EE{-6}&        1.61\EE{-6}&        1.17\EE{-6}&        8.82\EE{-7}\\
60&                 1.93\EE{-10}&       1.70\EE{-10}&       2.09\EE{-11}&       5.12\EE{-4}&        1.07\EE{-9}&        1.51\EE{-9}&        3.28\EE{-9}&        2.78\EE{-10}&       7.52\EE{-13}&       2.96\EE{-13}&       2.36\EE{-13}&       1.88\EE{-13}&       5.71\EE{-13}&       4.68\EE{-13}&       3.86\EE{-13}\\
62&                 8.30\EE{-6}&        1.94\EE{-5}&        1.09\EE{-5}&        6.60\EE{-5}&        3.76\EE{-3}&        6.16\EE{-3}&        1.74\EE{-2}&        5.94\EE{-4}&        1.29\EE{-4}&        5.62\EE{-5}&        3.17\EE{-5}&        2.03\EE{-5}&        1.39\EE{-5}&        1.00\EE{-5}&        7.51\EE{-6}\\

\end{longtable}


\clearpage

{\scriptsize

\begin{longtable}{l@{ }l@{ }l@{ }l@{ }l@{ }l@{ }l@{ }l@{ }l@{ }l@{ }l@{ }l@{ }l@{ }l@{ }l@{ }l@{ }l@{ }l}
\caption{Free-bound $\OS$-values for bound symmetry $J^{\Par} = 3/2^{\rm{o}}$ obtained by
integrating \pcs s from \Rm-matrix calculations. The superscript denotes the power of 10
by which the number is to be multiplied. \label{fValues263o}}\\
\hline\hline %
In.&                \cC{2}&             \cC{13}&            \cC{14}&            \cC{18}&            \cC{21}&            \cC{23}&            \cC{28}&            \cC{33}&            \cC{35}&            \cC{50}&            \cC{65}&            \cC{78}&            \cC{89}&            \cC{100}&           \cC{111}&           \cC{122}&           \cC{134}\\
\hline %
\endfirsthead %
\caption[]{continued.}\\ %
\hline\hline %
In.&                \cC{2}&             \cC{13}&            \cC{14}&            \cC{18}&            \cC{21}&            \cC{23}&            \cC{28}&            \cC{33}&            \cC{35}&            \cC{50}&            \cC{65}&            \cC{78}&            \cC{89}&            \cC{100}&           \cC{111}&           \cC{122}&           \cC{134}\\
\hline %
\endhead %
\hline %
\endfoot %
\vspace{-0.2cm} \\
15&                 1.49\EE{-3}&        3.73\EE{-4}&        3.75\EE{-10}&       2.32\EE{-3}&        1.45\EE{-4}&        6.09\EE{-6}&        4.82\EE{-3}&        2.45\EE{-4}&        7.17\EE{-3}&        9.78\EE{-4}&        3.82\EE{-4}&        2.21\EE{-4}&        1.47\EE{-4}&        1.04\EE{-4}&        7.77\EE{-5}&        5.97\EE{-5}&        4.69\EE{-5}\\
16&                 7.53\EE{-3}&        1.91\EE{-3}&        1.58\EE{-9}&        4.40\EE{-4}&        7.03\EE{-4}&        2.70\EE{-5}&        2.39\EE{-2}&        1.23\EE{-3}&        3.60\EE{-2}&        4.91\EE{-3}&        1.92\EE{-3}&        1.11\EE{-3}&        7.36\EE{-4}&        5.25\EE{-4}&        3.91\EE{-4}&        2.99\EE{-4}&        2.35\EE{-4}\\
17&                 4.15\EE{-6}&        1.12\EE{-6}&        1.96\EE{-10}&       6.73\EE{-6}&        4.15\EE{-7}&        1.33\EE{-3}&        1.48\EE{-5}&        7.06\EE{-7}&        2.02\EE{-5}&        2.73\EE{-6}&        1.06\EE{-6}&        6.09\EE{-7}&        4.02\EE{-7}&        2.85\EE{-7}&        2.11\EE{-7}&        1.61\EE{-7}&        1.26\EE{-7}\\
18&                 9.38\EE{-6}&        2.33\EE{-6}&        1.26\EE{-9}&        2.57\EE{-7}&        9.49\EE{-7}&        8.54\EE{-3}&        3.96\EE{-5}&        1.87\EE{-6}&        5.00\EE{-5}&        6.55\EE{-6}&        2.40\EE{-6}&        1.38\EE{-6}&        9.09\EE{-7}&        6.43\EE{-7}&        4.53\EE{-7}&        3.45\EE{-7}&        2.68\EE{-7}\\
19&                 8.52\EE{-7}&        6.35\EE{-7}&        2.88\EE{-9}&        1.31\EE{-8}&        6.29\EE{-8}&        1.70\EE{-2}&        3.61\EE{-7}&        1.17\EE{-7}&        4.99\EE{-7}&        4.47\EE{-7}&        2.97\EE{-7}&        2.18\EE{-7}&        1.67\EE{-7}&        1.30\EE{-7}&        1.03\EE{-7}&        8.24\EE{-8}&        6.72\EE{-8}\\
21&                 1.07\EE{-6}&        3.31\EE{-7}&        9.71\EE{-9}&        7.28\EE{-8}&        1.41\EE{-7}&        1.73\EE{-3}&        4.13\EE{-6}&        1.77\EE{-7}&        5.05\EE{-6}&        6.70\EE{-7}&        2.56\EE{-7}&        1.45\EE{-7}&        9.47\EE{-8}&        6.65\EE{-8}&        4.88\EE{-8}&        3.70\EE{-8}&        2.88\EE{-8}\\
22&                 8.46\EE{-8}&        1.94\EE{-7}&        3.34\EE{-6}&        2.59\EE{-8}&        8.10\EE{-8}&        6.20\EE{-4}&        5.42\EE{-7}&        5.30\EE{-8}&        2.18\EE{-8}&        3.66\EE{-8}&        3.16\EE{-8}&        2.63\EE{-8}&        2.16\EE{-8}&        1.76\EE{-8}&        1.44\EE{-8}&        1.18\EE{-8}&        8.36\EE{-9}\\
23&                 8.05\EE{-7}&        3.16\EE{-7}&        6.69\EE{-6}&        3.12\EE{-7}&        7.82\EE{-8}&        1.75\EE{-4}&        8.71\EE{-7}&        5.77\EE{-8}&        1.69\EE{-6}&        3.10\EE{-7}&        1.41\EE{-7}&        8.84\EE{-8}&        6.13\EE{-8}&        4.48\EE{-8}&        3.38\EE{-8}&        2.61\EE{-8}&        2.06\EE{-8}\\
24&                 1.21\EE{-5}&        7.19\EE{-6}&        1.02\EE{-5}&        1.33\EE{-7}&        1.97\EE{-6}&        5.70\EE{-4}&        1.39\EE{-6}&        9.26\EE{-7}&        9.35\EE{-6}&        4.66\EE{-6}&        2.71\EE{-6}&        1.93\EE{-6}&        1.44\EE{-6}&        1.11\EE{-6}&        8.68\EE{-7}&        6.92\EE{-7}&        5.60\EE{-7}\\
25&                 2.17\EE{-3}&        1.34\EE{-3}&        4.07\EE{-10}&       2.32\EE{-3}&        3.60\EE{-4}&        1.03\EE{-7}&        2.82\EE{-4}&        1.55\EE{-4}&        1.81\EE{-3}&        7.25\EE{-4}&        4.35\EE{-4}&        3.06\EE{-4}&        2.27\EE{-4}&        1.74\EE{-4}&        1.36\EE{-4}&        1.08\EE{-4}&        8.72\EE{-5}\\
26&                 1.65\EE{-2}&        1.03\EE{-2}&        7.75\EE{-9}&        2.52\EE{-4}&        2.85\EE{-3}&        1.34\EE{-6}&        2.04\EE{-3}&        6.47\EE{-4}&        1.55\EE{-2}&        6.42\EE{-3}&        3.82\EE{-3}&        2.70\EE{-3}&        2.01\EE{-3}&        1.40\EE{-3}&        1.10\EE{-3}&        8.74\EE{-4}&        7.06\EE{-4}\\
27&                 1.83\EE{-3}&        2.69\EE{-3}&        5.75\EE{-9}&        3.73\EE{-8}&        6.92\EE{-4}&        9.90\EE{-8}&        1.46\EE{-3}&        4.33\EE{-4}&        2.26\EE{-3}&        4.40\EE{-4}&        2.77\EE{-4}&        2.29\EE{-4}&        1.76\EE{-4}&        1.39\EE{-4}&        1.11\EE{-4}&        8.97\EE{-5}&        7.34\EE{-5}\\
43&                 8.82\EE{-10}&       3.29\EE{-8}&        2.11\EE{-11}&       1.95\EE{-2}&        4.70\EE{-10}&       3.85\EE{-7}&        4.71\EE{-6}&        9.11\EE{-6}&        2.73\EE{-5}&        9.58\EE{-7}&        2.10\EE{-7}&        9.16\EE{-8}&        5.05\EE{-8}&        3.29\EE{-8}&        2.22\EE{-8}&        1.56\EE{-8}&        1.13\EE{-8}\\
45&                 1.40\EE{-10}&       3.21\EE{-9}&        1.02\EE{-13}&       1.56\EE{-6}&        1.28\EE{-9}&        4.30\EE{-7}&        3.43\EE{-7}&        5.82\EE{-7}&        1.77\EE{-6}&        6.67\EE{-8}&        1.43\EE{-8}&        7.03\EE{-9}&        4.33\EE{-9}&        3.03\EE{-9}&        2.30\EE{-9}&        1.85\EE{-9}&        1.55\EE{-9}\\
46&                 3.92\EE{-10}&       1.91\EE{-8}&        3.10\EE{-12}&       8.73\EE{-4}&        9.29\EE{-9}&        1.59\EE{-6}&        3.18\EE{-6}&        5.65\EE{-6}&        1.69\EE{-5}&        5.77\EE{-7}&        1.32\EE{-7}&        6.04\EE{-8}&        3.58\EE{-8}&        2.41\EE{-8}&        1.77\EE{-8}&        1.37\EE{-8}&        1.11\EE{-8}\\
51&                 2.59\EE{-10}&       2.29\EE{-9}&        4.10\EE{-13}&       2.84\EE{-5}&        1.02\EE{-9}&        2.51\EE{-8}&        2.90\EE{-7}&        4.87\EE{-7}&        1.41\EE{-6}&        4.77\EE{-8}&        1.03\EE{-8}&        4.54\EE{-9}&        2.56\EE{-9}&        1.64\EE{-9}&        1.13\EE{-9}&        8.19\EE{-10}&       6.15\EE{-10}\\
58&                 2.89\EE{-6}&        9.08\EE{-6}&        8.44\EE{-10}&       3.22\EE{-5}&        3.81\EE{-6}&        5.82\EE{-4}&        1.25\EE{-3}&        2.06\EE{-3}&        5.93\EE{-3}&        1.97\EE{-4}&        4.20\EE{-5}&        1.81\EE{-5}&        1.05\EE{-5}&        6.70\EE{-6}&        4.61\EE{-6}&        3.32\EE{-6}&        2.49\EE{-6}\\
59&                 7.73\EE{-9}&        3.06\EE{-7}&        2.55\EE{-11}&       1.70\EE{-4}&        1.65\EE{-7}&        3.98\EE{-4}&        4.12\EE{-5}&        6.90\EE{-5}&        1.98\EE{-4}&        6.91\EE{-6}&        1.45\EE{-6}&        6.29\EE{-7}&        3.54\EE{-7}&        2.39\EE{-7}&        1.64\EE{-7}&        1.19\EE{-7}&        8.89\EE{-8}\\
60&                 1.58\EE{-11}&       1.07\EE{-12}&       1.03\EE{-10}&       3.51\EE{-11}&       9.86\EE{-12}&       5.14\EE{-5}&        2.41\EE{-10}&       3.99\EE{-10}&       8.16\EE{-10}&       1.11\EE{-10}&       6.77\EE{-13}&       5.32\EE{-13}&       4.18\EE{-13}&       3.31\EE{-13}&       3.03\EE{-13}&       2.46\EE{-13}&       2.00\EE{-13}\\
61&                 6.45\EE{-6}&        1.47\EE{-5}&        5.74\EE{-9}&        2.85\EE{-5}&        7.05\EE{-6}&        2.76\EE{-4}&        2.50\EE{-3}&        4.03\EE{-3}&        1.17\EE{-2}&        4.03\EE{-4}&        8.49\EE{-5}&        3.67\EE{-5}&        2.07\EE{-5}&        1.31\EE{-5}&        8.95\EE{-6}&        6.41\EE{-6}&        4.77\EE{-6}\\
62&                 6.54\EE{-7}&        1.93\EE{-6}&        8.56\EE{-10}&       1.53\EE{-3}&        1.60\EE{-6}&        4.23\EE{-5}&        3.67\EE{-4}&        5.87\EE{-4}&        1.70\EE{-3}&        6.05\EE{-5}&        1.31\EE{-5}&        5.72\EE{-6}&        3.24\EE{-6}&        2.07\EE{-6}&        1.42\EE{-6}&        1.02\EE{-6}&        7.67\EE{-7}\\

\end{longtable}

}


\clearpage

\begin{longtable}{@{\extracolsep\fill}llllllllllllr@{}}
\caption{Free-bound $\OS$-values for bound symmetry $J^{\Par} = 5/2^{\rm{o}}$ obtained by
integrating \pcs s from \Rm-matrix calculations. The superscript denotes the power of 10
by which the number is to be multiplied. \label{fValues265o}}\\
\hline\hline %
In.&                \cC{17}&            \cC{24}&            \cC{29}&            \cC{38}&            \cC{53}&            \cC{68}&            \cC{79}&            \cC{92}&            \cC{103}&           \cC{114}&           \cC{126}&           \cC{137}\\
\hline %
\endfirsthead %
\caption[]{continued.}\\ %
\hline\hline %
In.&                \cC{17}&            \cC{24}&            \cC{29}&            \cC{38}&            \cC{53}&            \cC{68}&            \cC{79}&            \cC{92}&            \cC{103}&           \cC{114}&           \cC{126}&           \cC{137}\\
\hline %
\endhead %
\hline %
\endfoot %
\vspace{-0.2cm} \\
16&                 2.78\EE{-3}&        2.11\EE{-7}&        1.34\EE{-8}&        8.88\EE{-9}&        8.79\EE{-9}&        8.27\EE{-9}&        7.47\EE{-9}&        6.57\EE{-9}&        5.69\EE{-9}&        4.90\EE{-9}&        4.15\EE{-9}&        3.61\EE{-9}\\
18&                 4.47\EE{-6}&        5.27\EE{-4}&        3.69\EE{-9}&        3.39\EE{-9}&        3.30\EE{-9}&        3.25\EE{-9}&        3.14\EE{-9}&        3.18\EE{-9}&        3.12\EE{-9}&        3.09\EE{-9}&        2.98\EE{-9}&        3.21\EE{-9}\\
19&                 1.24\EE{-7}&        4.78\EE{-3}&        4.64\EE{-10}&       4.43\EE{-10}&       4.37\EE{-10}&       4.30\EE{-10}&       4.17\EE{-10}&       4.00\EE{-10}&       3.82\EE{-10}&       3.48\EE{-10}&       3.66\EE{-10}&       3.99\EE{-10}\\
20&                 2.32\EE{-9}&        1.20\EE{-2}&        1.72\EE{-17}&       1.38\EE{-16}&       3.10\EE{-15}&       0&                  0&                  0&                  0&                  0&                  0&                  0\\
21&                 4.56\EE{-7}&        1.61\EE{-3}&        3.64\EE{-13}&       3.65\EE{-13}&       3.21\EE{-13}&       1.59\EE{-13}&       1.15\EE{-13}&       4.49\EE{-15}&       7.60\EE{-14}&       7.09\EE{-14}&       7.56\EE{-14}&       1.44\EE{-13}\\
23&                 2.24\EE{-8}&        5.24\EE{-4}&        9.47\EE{-9}&        8.94\EE{-9}&        9.08\EE{-9}&        8.40\EE{-9}&        6.36\EE{-9}&        5.54\EE{-9}&        4.76\EE{-9}&        4.07\EE{-9}&        3.93\EE{-9}&        3.43\EE{-9}\\
24&                 1.24\EE{-6}&        1.07\EE{-3}&        3.20\EE{-9}&        2.91\EE{-9}&        2.67\EE{-9}&        2.43\EE{-9}&        2.17\EE{-9}&        1.92\EE{-9}&        1.69\EE{-9}&        1.48\EE{-9}&        1.30\EE{-9}&        1.14\EE{-9}\\
25&                 1.57\EE{-4}&        1.52\EE{-10}&       5.59\EE{-5}&        4.91\EE{-5}&        3.89\EE{-5}&        3.48\EE{-5}&        3.05\EE{-5}&        2.63\EE{-5}&        2.26\EE{-5}&        1.93\EE{-5}&        1.65\EE{-5}&        1.25\EE{-5}\\
26&                 2.29\EE{-3}&        3.79\EE{-8}&        3.50\EE{-6}&        3.16\EE{-6}&        2.83\EE{-6}&        2.51\EE{-6}&        2.20\EE{-6}&        1.90\EE{-6}&        1.63\EE{-6}&        1.39\EE{-6}&        1.19\EE{-6}&        9.00\EE{-7}\\
43&                 8.17\EE{-4}&        1.17\EE{-7}&        3.26\EE{-7}&        5.10\EE{-7}&        6.46\EE{-7}&        7.25\EE{-7}&        7.59\EE{-7}&        7.58\EE{-7}&        7.35\EE{-7}&        6.97\EE{-7}&        6.51\EE{-7}&        6.02\EE{-7}\\
44&                 1.46\EE{-2}&        2.83\EE{-6}&        9.13\EE{-8}&        7.43\EE{-8}&        9.39\EE{-8}&        1.06\EE{-7}&        1.12\EE{-7}&        1.14\EE{-7}&        1.12\EE{-7}&        1.08\EE{-7}&        1.03\EE{-7}&        9.68\EE{-8}\\
45&                 1.09\EE{-7}&        2.81\EE{-8}&        1.63\EE{-8}&        2.04\EE{-10}&       2.08\EE{-10}&       2.01\EE{-10}&       1.94\EE{-10}&       1.87\EE{-10}&       1.83\EE{-10}&       1.98\EE{-10}&       2.20\EE{-10}&       2.44\EE{-10}\\
46&                 2.49\EE{-5}&        4.66\EE{-7}&        3.70\EE{-9}&        2.17\EE{-8}&        2.74\EE{-8}&        3.06\EE{-8}&        3.17\EE{-8}&        3.13\EE{-8}&        2.99\EE{-8}&        2.77\EE{-8}&        2.50\EE{-8}&        2.22\EE{-8}\\
47&                 4.70\EE{-3}&        9.64\EE{-7}&        4.28\EE{-8}&        9.19\EE{-9}&        1.03\EE{-8}&        1.06\EE{-8}&        1.05\EE{-8}&        9.61\EE{-9}&        8.45\EE{-9}&        7.13\EE{-9}&        5.78\EE{-9}&        4.46\EE{-9}\\
51&                 5.49\EE{-7}&        7.39\EE{-9}&        1.91\EE{-8}&        2.24\EE{-10}&       2.40\EE{-10}&       2.20\EE{-10}&       1.69\EE{-10}&       1.04\EE{-10}&       4.31\EE{-11}&       4.57\EE{-12}&       1.03\EE{-11}&       9.47\EE{-11}\\
52&                 3.19\EE{-7}&        1.35\EE{-7}&        2.87\EE{-7}&        7.10\EE{-7}&        9.12\EE{-7}&        1.04\EE{-6}&        1.11\EE{-6}&        1.13\EE{-6}&        1.12\EE{-6}&        1.09\EE{-6}&        1.04\EE{-6}&        9.99\EE{-7}\\
55&                 5.58\EE{-5}&        6.96\EE{-8}&        3.70\EE{-6}&        5.17\EE{-6}&        6.57\EE{-6}&        7.40\EE{-6}&        7.76\EE{-6}&        7.76\EE{-6}&        7.51\EE{-6}&        7.11\EE{-6}&        6.62\EE{-6}&        6.08\EE{-6}\\
57&                 1.23\EE{-5}&        9.69\EE{-4}&        1.91\EE{-9}&        3.11\EE{-10}&       4.07\EE{-10}&       4.68\EE{-10}&       5.02\EE{-10}&       5.15\EE{-10}&       5.15\EE{-10}&       5.06\EE{-10}&       4.92\EE{-10}&       4.75\EE{-10}\\
58&                 6.08\EE{-4}&        1.62\EE{-4}&        1.33\EE{-8}&        5.77\EE{-9}&        5.26\EE{-9}&        4.56\EE{-9}&        3.83\EE{-9}&        3.16\EE{-9}&        2.57\EE{-9}&        2.10\EE{-9}&        1.70\EE{-9}&        1.41\EE{-9}\\
59&                 1.27\EE{-5}&        2.48\EE{-5}&        8.56\EE{-9}&        2.70\EE{-8}&        2.53\EE{-8}&        2.30\EE{-8}&        2.04\EE{-8}&        1.77\EE{-8}&        1.52\EE{-8}&        1.31\EE{-8}&        1.11\EE{-8}&        9.56\EE{-9}\\
61&                 1.26\EE{-3}&        8.08\EE{-5}&        1.74\EE{-9}&        4.53\EE{-9}&        3.89\EE{-9}&        3.29\EE{-9}&        2.76\EE{-9}&        2.31\EE{-9}&        1.95\EE{-9}&        1.65\EE{-9}&        1.41\EE{-9}&        1.24\EE{-9}\\
62&                 1.12\EE{-4}&        2.75\EE{-6}&        2.04\EE{-7}&        2.38\EE{-7}&        2.25\EE{-7}&        2.05\EE{-7}&        1.81\EE{-7}&        1.57\EE{-7}&        1.36\EE{-7}&        1.17\EE{-7}&        9.93\EE{-8}&        8.59\EE{-8}\\

\end{longtable}


\clearpage

{\small
\begin{longtable}{@{\extracolsep\fill}llllllllll@{}}
\caption{Sample of our line list where several lines have been observed astronomically. The first
column is for experimental/theoretical energy identification for the upper and lower states
respectively where 1 refers to experimental while 0 refers to theoretical. The other columns are
for the atomic designation of the upper and lower states respectively, followed by the air
wavelength in angstrom and effective dielectronic recombination rate coefficients in
cm$^3$.s$^{-1}$ for the given logarithmic temperatures. The superscript denotes the power of 10 by
which the number is to be multiplied. \label{ListSample}}
    {\bf } &     {\bf } &     {\bf } &     {\bf } &                                           \multicolumn{ 6}{c}{{\bf log10(T)}} \\

  {\bf ET} & {\bf Upper} & {\bf Lower} & {\bf $\WL_{air}$} &  {\bf 2.0} &  {\bf 2.5} &  {\bf 3.0} &  {\bf 3.5} &  {\bf 4.0} &  {\bf 4.5} \\
\hline
        11 & 2s2p(\SLP3Po)4f \SLPJ4De{1/2} & 2s2p(\SLP3Po)3d \SLPJ2Po{1/2} &    5116.73 &  5.53\EE{-190} &   2.59\EE{-81} &   1.85\EE{-47} &   2.89\EE{-37} &   1.48\EE{-34} &   3.27\EE{-34} \\

        11 & 2s2p(\SLP3Po)4f \SLPJ4De{5/2} & 2s2p(\SLP3Po)3d \SLPJ2Po{3/2} &    5119.46 &  5.14\EE{-168} &   9.10\EE{-62} &   1.11\EE{-28} &   9.88\EE{-19} &   4.25\EE{-16} &   8.88\EE{-16} \\

        11 & 2s2p(\SLP3Po)3p \SLPJ2Pe{3/2} & 2s$^2$4p \SLPJ2Po{1/2} &    5120.08 &   2.50\EE{-26} &   1.64\EE{-19} &   7.84\EE{-18} &   3.26\EE{-17} &   3.79\EE{-17} &   1.76\EE{-17} \\

        11 & 2s2p(\SLP3Po)4f \SLPJ4De{3/2} & 2s2p(\SLP3Po)3d \SLPJ2Po{1/2} &    5120.17 &  6.88\EE{-169} &   1.51\EE{-62} &   1.97\EE{-29} &   1.80\EE{-19} &   7.78\EE{-17} &   1.63\EE{-16} \\

        11 & 2s2p(\SLP3Po)3p \SLPJ2Pe{3/2} & 2s$^2$4p \SLPJ2Po{3/2} &    5121.83 &   1.20\EE{-25} &   7.87\EE{-19} &   3.75\EE{-17} &   1.56\EE{-16} &   1.81\EE{-16} &   8.40\EE{-17} \\

        11 & 2s$^2$7f \SLPJ2Fo{5/2} & 2s$^2$4d \SLPJ2De{3/2} &    5122.09 &   6.05\EE{-29} &   7.95\EE{-22} &   2.03\EE{-18} &   9.32\EE{-18} &   6.66\EE{-18} &   3.55\EE{-18} \\

        11 & 2s$^2$7f \SLPJ2Fo{7/2} & 2s$^2$4d \SLPJ2De{5/2} &    5122.27 &   2.11\EE{-28} &   1.95\EE{-21} &   3.05\EE{-18} &   1.44\EE{-17} &   1.03\EE{-17} &   5.43\EE{-18} \\

        11 & 2s$^2$7f \SLPJ2Fo{5/2} & 2s$^2$4d \SLPJ2De{5/2} &    5122.27 &   4.32\EE{-30} &   5.68\EE{-23} &   1.45\EE{-19} &   6.66\EE{-19} &   4.76\EE{-19} &   2.54\EE{-19} \\

        11 & 2s2p(\SLP3Po)3p \SLPJ2Pe{1/2} & 2s$^2$4p \SLPJ2Po{1/2} &    5125.21 &   1.46\EE{-26} &   8.64\EE{-20} &   4.82\EE{-18} &   5.15\EE{-17} &   7.05\EE{-17} &   3.50\EE{-17} \\

        11 & 2s2p(\SLP3Po)3p \SLPJ2Pe{1/2} & 2s$^2$4p \SLPJ2Po{3/2} &    5126.96 &   7.08\EE{-27} &   4.20\EE{-20} &   2.34\EE{-18} &   2.50\EE{-17} &   3.42\EE{-17} &   1.70\EE{-17} \\

        11 & 2s2p(\SLP3Po)3p \SLPJ4Pe{3/2} & 2s2p(\SLP3Po)3s \SLPJ4Po{1/2} &    5132.95 &   4.96\EE{-28} &   9.80\EE{-20} &   1.65\EE{-17} &   2.49\EE{-16} &   2.26\EE{-15} &   1.48\EE{-15} \\

        11 & 2s2p(\SLP3Po)3p \SLPJ4Pe{5/2} & 2s2p(\SLP3Po)3s \SLPJ4Po{3/2} &    5133.28 &   1.69\EE{-27} &   1.09\EE{-19} &   1.71\EE{-17} &   1.66\EE{-16} &   1.50\EE{-15} &   1.08\EE{-15} \\

        11 & 2s2p(\SLP3Po)3p \SLPJ4Pe{1/2} & 2s2p(\SLP3Po)3s \SLPJ4Po{1/2} &    5137.26 &   1.15\EE{-28} &   5.07\EE{-20} &   8.50\EE{-18} &   9.92\EE{-17} &   8.62\EE{-16} &   5.55\EE{-16} \\

        11 & 2s2p(\SLP3Po)3p \SLPJ4Pe{3/2} & 2s2p(\SLP3Po)3s \SLPJ4Po{3/2} &    5139.17 &   1.57\EE{-28} &   3.10\EE{-20} &   5.23\EE{-18} &   7.89\EE{-17} &   7.15\EE{-16} &   4.68\EE{-16} \\

        01 & 2s$^2$11p \SLPJ2Po{3/2} & 2s$^2$5s \SLPJ2Se{1/2} &    5141.70 &   9.29\EE{-33} &   1.36\EE{-25} &   3.26\EE{-21} &   1.25\EE{-19} &   6.35\EE{-19} &   6.92\EE{-19} \\

        01 & 2s$^2$11p \SLPJ2Po{1/2} & 2s$^2$5s \SLPJ2Se{1/2} &    5141.76 &   2.10\EE{-33} &   4.79\EE{-26} &   1.60\EE{-21} &   6.26\EE{-20} &   3.16\EE{-19} &   3.48\EE{-19} \\

        11 & 2s2p(\SLP3Po)3p \SLPJ4Pe{1/2} & 2s2p(\SLP3Po)3s \SLPJ4Po{3/2} &    5143.49 &   5.77\EE{-28} &   2.55\EE{-19} &   4.27\EE{-17} &   4.99\EE{-16} &   4.34\EE{-15} &   2.79\EE{-15} \\

        11 & 2s2p(\SLP3Po)3p \SLPJ4Pe{5/2} & 2s2p(\SLP3Po)3s \SLPJ4Po{5/2} &    5145.16 &   3.97\EE{-27} &   2.56\EE{-19} &   4.01\EE{-17} &   3.90\EE{-16} &   3.54\EE{-15} &   2.53\EE{-15} \\

        11 & 2s2p(\SLP3Po)3p \SLPJ4Pe{3/2} & 2s2p(\SLP3Po)3s \SLPJ4Po{5/2} &    5151.08 &   5.59\EE{-28} &   1.11\EE{-19} &   1.86\EE{-17} &   2.81\EE{-16} &   2.55\EE{-15} &   1.67\EE{-15} \\

        11 & 2s2p(\SLP3Po)4f \SLPJ4Ge{5/2} & 2s2p(\SLP3Po)3d \SLPJ2Po{3/2} &    5162.53 &  1.92\EE{-171} &   6.83\EE{-66} &   5.01\EE{-33} &   3.80\EE{-23} &   1.56\EE{-20} &   3.20\EE{-20} \\

        11 & 2s2p(\SLP3Po)3d \SLPJ2Do{5/2} & 2s$^2$5g \SLPJ2Ge{7/2} &    5162.98 &   7.68\EE{-33} &   5.07\EE{-26} &   2.23\EE{-24} &   2.27\EE{-24} &   7.55\EE{-25} &   2.38\EE{-25} \\

        00 & 2s2p(\SLP3Po)4s \SLPJ2Po{3/2} & 2s$^2$11s \SLPJ2Se{1/2} &    5172.37 &  8.25\EE{-108} &   1.12\EE{-43} &   6.57\EE{-24} &   3.60\EE{-18} &   7.20\EE{-17} &   5.71\EE{-17} \\

        00 & 2s2p(\SLP3Po)4s \SLPJ2Po{1/2} & 2s$^2$11s \SLPJ2Se{1/2} &    5187.14 &  9.10\EE{-108} &   7.19\EE{-44} &   3.55\EE{-24} &   1.84\EE{-18} &   3.63\EE{-17} &   2.86\EE{-17} \\

        10 & 2s2p(\SLP3Po)4s \SLPJ4Po{3/2} & 2s$^2$9s \SLPJ2Se{1/2} &    5196.92 &   7.54\EE{-98} &   1.97\EE{-43} &   9.74\EE{-27} &   5.69\EE{-22} &   5.63\EE{-21} &   3.57\EE{-21} \\

        10 & 2s2p(\SLP3Po)4s \SLPJ4Po{1/2} & 2s$^2$9s \SLPJ2Se{1/2} &    5203.44 &   1.74\EE{-98} &   3.58\EE{-44} &   1.65\EE{-27} &   9.41\EE{-23} &   9.23\EE{-22} &   5.84\EE{-22} \\

        11 & 2s2p(\SLP3Po)3p \SLPJ4De{5/2} & 2s$^2$4p \SLPJ2Po{3/2} &    5203.77 &   7.63\EE{-33} &   2.44\EE{-24} &   4.07\EE{-22} &   7.12\EE{-21} &   1.35\EE{-19} &   1.71\EE{-19} \\

        11 & 2s2p(\SLP3Po)4p \SLPJ4Se{3/2} & 2s2p(\SLP3Po)3d \SLPJ4Do{1/2} &    5204.20 &  2.74\EE{-137} &   2.03\EE{-56} &   2.33\EE{-31} &   6.55\EE{-24} &   5.86\EE{-22} &   8.17\EE{-22} \\

        11 & 2s2p(\SLP3Po)4p \SLPJ4Se{3/2} & 2s2p(\SLP3Po)3d \SLPJ4Do{3/2} &    5205.70 &  1.73\EE{-136} &   1.28\EE{-55} &   1.48\EE{-30} &   4.15\EE{-23} &   3.71\EE{-21} &   5.18\EE{-21} \\

        11 & 2s2p(\SLP3Po)4p \SLPJ4Se{3/2} & 2s2p(\SLP3Po)3d \SLPJ4Do{5/2} &    5207.98 &  3.34\EE{-136} &   2.47\EE{-55} &   2.84\EE{-30} &   8.00\EE{-23} &   7.15\EE{-21} &   9.98\EE{-21} \\

        11 & 2s2p(\SLP3Po)3p \SLPJ4De{3/2} & 2s$^2$4p \SLPJ2Po{1/2} &    5208.74 &   1.87\EE{-31} &   3.80\EE{-24} &   4.99\EE{-22} &   1.05\EE{-20} &   1.62\EE{-19} &   1.83\EE{-19} \\

        11 & 2s2p(\SLP3Po)3p \SLPJ4De{3/2} & 2s$^2$4p \SLPJ2Po{3/2} &    5210.56 &   2.50\EE{-30} &   5.10\EE{-23} &   6.70\EE{-21} &   1.41\EE{-19} &   2.17\EE{-18} &   2.46\EE{-18} \\

        11 & 2s2p(\SLP3Po)3p \SLPJ4De{1/2} & 2s$^2$4p \SLPJ2Po{1/2} &    5212.65 &   1.15\EE{-30} &   2.43\EE{-23} &   3.42\EE{-21} &   1.01\EE{-19} &   1.06\EE{-18} &   8.41\EE{-19} \\

        11 & 2s2p(\SLP3Po)3p \SLPJ4De{1/2} & 2s$^2$4p \SLPJ2Po{3/2} &    5214.46 &   5.52\EE{-31} &   1.17\EE{-23} &   1.64\EE{-21} &   4.85\EE{-20} &   5.09\EE{-19} &   4.03\EE{-19} \\

        11 & 2s2p(\SLP3Po)4p \SLPJ4De{7/2} & 2s2p(\SLP3Po)3d \SLPJ4Fo{5/2} &    5244.05 &  1.17\EE{-158} &   1.89\EE{-61} &   3.48\EE{-31} &   4.16\EE{-22} &   9.58\EE{-20} &   1.65\EE{-19} \\

        11 & 2s2p(\SLP3Po)4p \SLPJ4De{7/2} & 2s2p(\SLP3Po)3d \SLPJ4Fo{7/2} &    5249.53 &  2.53\EE{-157} &   4.08\EE{-60} &   7.50\EE{-30} &   8.98\EE{-21} &   2.07\EE{-18} &   3.55\EE{-18} \\

        11 & 2s2p(\SLP3Po)4p \SLPJ4De{5/2} & 2s2p(\SLP3Po)3d \SLPJ4Fo{3/2} &    5249.90 &  1.29\EE{-126} &   6.68\EE{-50} &   3.73\EE{-26} &   3.70\EE{-19} &   1.86\EE{-17} &   1.97\EE{-17} \\

        11 & 2s2p(\SLP3Po)4p \SLPJ4De{5/2} & 2s2p(\SLP3Po)3d \SLPJ4Fo{5/2} &    5253.56 &  2.49\EE{-125} &   1.29\EE{-48} &   7.20\EE{-25} &   7.15\EE{-18} &   3.59\EE{-16} &   3.80\EE{-16} \\

        11 & 2s2p(\SLP3Po)4p \SLPJ4De{3/2} & 2s2p(\SLP3Po)3d \SLPJ4Fo{3/2} &    5256.09 &  2.63\EE{-125} &   1.09\EE{-48} &   5.69\EE{-25} &   5.52\EE{-18} &   2.75\EE{-16} &   2.91\EE{-16} \\

        11 & 2s2p(\SLP3Po)4p \SLPJ4De{7/2} & 2s2p(\SLP3Po)3d \SLPJ4Fo{9/2} &    5257.24 &  2.45\EE{-156} &   3.96\EE{-59} &   7.27\EE{-29} &   8.71\EE{-20} &   2.01\EE{-17} &   3.44\EE{-17} \\

        11 & 2s2p(\SLP3Po)4p \SLPJ4De{5/2} & 2s2p(\SLP3Po)3d \SLPJ4Fo{7/2} &    5259.06 &  1.26\EE{-124} &   6.53\EE{-48} &   3.64\EE{-24} &   3.61\EE{-17} &   1.81\EE{-15} &   1.92\EE{-15} \\

        11 & 2s2p(\SLP3Po)4p \SLPJ4De{1/2} & 2s2p(\SLP3Po)3d \SLPJ4Fo{3/2} &    5259.66 &  1.46\EE{-126} &   5.35\EE{-50} &   2.67\EE{-26} &   2.65\EE{-19} &   1.51\EE{-17} &   1.78\EE{-17} \\
\hline
\end{longtable}
}

\end{document}